\documentclass[12pt]{article}
\newlength{\pubnumber} 

\catcode`\@=11
\@addtoreset{equation}{section}

\def\section{\@startsection{section}{1}{\z@}{3.5ex plus 1ex minus .2ex}
 {2.3ex plus .2ex}{\large\bf}}
\def\subsection{\@startsection{subsection}{2}{\z@}{2.3ex plus .2ex}
 {2.3ex plus .2ex}{\bf}}

\begin{document}

\begin{titlepage}
\samepage{
\setcounter{page}{0}
\rightline{\tt hep-th/0312217}
\rightline{December 2003}
\vfill
\begin{center}
   {\Large \it  Adventures in Thermal Duality (II):\\}
    {\Large \bf Towards a Duality-Covariant String Thermodynamics\\ }
\vfill
   {\large
      Keith R. Dienes\footnote{
     E-mail address:  dienes@physics.arizona.edu}
        $\,$ and $\,$ Michael Lennek\footnote{
     E-mail address:  mlennek@physics.arizona.edu}
    \\}
\vspace{.12in}
 {\it  Department of Physics, University of Arizona, Tucson, AZ  85721  USA\\}
\end{center}
\vfill
\begin{abstract}
  {\rm 
   In a recent companion paper, we observed that the rules of ordinary
   thermodynamics generally fail to respect thermal duality,
   a symmetry of string theory under which the physics at temperature
   $T$ is related to the physics at the inverse temperature $1/T$. 
   Even when the free energy and internal energy exhibit the thermal
   duality symmetry, the entropy and specific heat are defined
   in such a way that this symmetry is destroyed.
   In this paper, we propose a modification of the traditional
   definitions of these quantities, yielding
   a manifestly duality-covariant thermodynamics.
   At low temperatures, these modifications produce ``corrections'' 
   to the standard
   definitions of entropy and specific heat 
   which are suppressed by powers of the string scale.  These
   corrections  
   may nevertheless be important for the full     
   development of a consistent string thermodynamics.
   We find, for example, that the string-corrected entropy
   can be smaller than the usual entropy at high temperatures, 
    suggesting a possible connection with the holographic principle.
   We also discuss some outstanding theoretical issues prompted by 
    our approach.
   }
\end{abstract}
\vfill
\smallskip}
\end{titlepage}

\setcounter{footnote}{0}

% ========================= DEFINITIONS ===================================
\def\beq{\begin{equation}}
\def\eeq{\end{equation}}
\def\beqn{\begin{eqnarray}}
\def\eeqn{\end{eqnarray}}
\def\half{{\textstyle{1\over 2}}}

\def\calO{{\cal O}}
\def\calE{{\cal E}}
\def\calT{{\cal T}}
\def\calM{{\cal M}}
\def\calF{{\cal F}}
\def\calY{{\cal Y}}
\def\calV{{\cal V}}
\def\rep#1{{\bf {#1}}}
\def\ie{{\it i.e.}\/}
\def\eg{{\it e.g.}\/}

\newcommand{\newc}{\newcommand}
\newc{\gsim}{\lower.7ex\hbox{$\;\stackrel{\textstyle>}{\sim}\;$}}
\newc{\lsim}{\lower.7ex\hbox{$\;\stackrel{\textstyle<}{\sim}\;$}}

%==============================================================================
\hyphenation{su-per-sym-met-ric non-su-per-sym-met-ric}
\hyphenation{space-time-super-sym-met-ric}
\hyphenation{mod-u-lar mod-u-lar--in-var-i-ant}
%==============================================================================

%================== BLACKBOARD BOLD CHARACTERS ==============================

\def\inbar{\,\vrule height1.5ex width.4pt depth0pt}

\def\IC{\relax\hbox{$\inbar\kern-.3em{\rm C}$}}
\def\IQ{\relax\hbox{$\inbar\kern-.3em{\rm Q}$}}
\def\IR{\relax{\rm I\kern-.18em R}}
 \font\cmss=cmss10 \font\cmsss=cmss10 at 7pt
\def\IZ{\relax\ifmmode\mathchoice
 {\hbox{\cmss Z\kern-.4em Z}}{\hbox{\cmss Z\kern-.4em Z}}
 {\lower.9pt\hbox{\cmsss Z\kern-.4em Z}}
 {\lower1.2pt\hbox{\cmsss Z\kern-.4em Z}}\else{\cmss Z\kern-.4em Z}\fi}

% Redefine caption to put text and formulas in smaller font
\long\def\@caption#1[#2]#3{\par\addcontentsline{\csname
  ext@#1\endcsname}{#1}{\protect\numberline{\csname
  the#1\endcsname}{\ignorespaces #2}}\begingroup
    \small
    \@parboxrestore
    \@makecaption{\csname fnum@#1\endcsname}{\ignorespaces #3}\par
  \endgroup}
\catcode`@=12

\input epsf
%============================== TEXT BEGINS HERE ============================

%=============================================================================
\section{Introduction}
\setcounter{footnote}{0}

In a recent companion paper~\cite{I}, we observed that the
rules of ordinary thermodynamics generally fail to respect
the thermal duality symmetry of string theory under which
the physics at temperature $T$ is related 
to the physics at temperature 
$T_c^2/T$, where $T_c$ is a critical (or self-dual) temperature related
to the string scale.
The reason for this failure is simple:
even though the string vacuum amplitude $\calV(T)$ might exhibit 
an invariance under this symmetry, 
with             $\calV(T) = \calV(T_c^2/T)$,
the subsequent temperature derivatives $d/dT$ that are needed in order
to calculate other thermodynamic quantities
generally destroy this symmetry.  This then results in 
quantities such as entropy and specific heat which fail
to exhibit thermal duality symmetries.

It is, of course, entirely possible that thermal duality
should be viewed only as an ``accidental'' symmetry 
of the string vacuum amplitude;  we thus 
would have no problem with the loss of this symmetry
when calculating other thermodynamic quantities.
However, given that the thermal duality symmetry of string
theory follows immediately from T-duality
and Lorentz invariance, this symmetry appears to be every bit as
deep as the dualities that occur at zero temperature.
Thus, it seems more natural to consider thermal duality 
as a fundamental property of a consistent string theory,
and demand that this symmetry hold for {\it all}\/ physically
relevant thermodynamic quantities.

This problem was considered in Ref.~\cite{I}, where it was shown that 
there exist certain special vacuum amplitudes $\calV(T)$ for which all thermodynamic
quantities exhibit the thermal duality symmetry. 
Moreover, it was shown that these solutions for $\calV(T)$ 
correspond to highly symmetric string modular integrals in which
the time/temperature direction is compactified
on $S^1$ (a circle) or $S^1/\IZ_2$ (a line segment).
Thus, for such constructions, there is no loss of thermal duality.  
In fact, as discussed in Ref.~\cite{I}, the constraint of thermal duality
in such cases may be of sufficient
strength to enable an exact, closed-form evaluation of the 
relevant thermodynamic quantities.

However, this method of restoring thermal duality 
is less than satisfactory.
Because this approach applies only for certain selected ground states,
it lacks the generality that should apply to a fundamental symmetry.
If thermal duality is to be considered an intrinsic property of
finite-temperature string theory (akin to T-duality), then the formulation
of the theory itself --- including its rules of calculation ---
should respect this symmetry regardless of the specific ground state. 

This argument should apply even if the specific string ground state 
in question
does not exhibit thermal duality (such as may occur in finite-temperature
string constructions utilizing
temperature-dependent Wilson lines).  Indeed, even when 
thermal duality is ``spontaneously broken'' in this way,
the theoretical definitions of all relevant physical thermodynamic quantities 
should still reflect this duality symmetry.
After all, it is certainly acceptable if the 
entropy or specific heat
fail to exhibit thermal duality because the ground state
fails to yield a 
duality-symmetric 
vacuum amplitude $\calV(T)$.   However, it is not acceptable if
this failure arises because the {\it definitions}\/ of the entropy or
specific heat in terms of $\calV(T)$ are themselves not duality covariant.
  
For this reason, we are motivated to develop an
alternative, fully covariant string thermodynamics
in which thermal duality is manifest.
This is the goal of the present paper (a short outline of our
basic ideas also appears in Ref.~\cite{summary}).
As we shall see,
this new duality-covariant framework 
will preserve the definitions of free energy and internal energy,
but will lead naturally to modifications in the usual thermodynamic
definitions for other quantities such as entropy and specific heat.  
At low temperatures, these modifications produce ``corrections'' 
to the standard
definitions of entropy and specific heat 
which are suppressed by powers of the string scale.  
These corrections are therefore 
unanticipated from the low-energy (low-temperature) point of view.
At higher temperatures, however, these modifications are significant,
and may be important for a full understanding of 
string thermodynamics at or near the self-dual temperature $T_c$.
In fact, we shall find that our new, string-corrected entropy
is often smaller than the usual entropy, with the suppression
becoming increasingly severe as the temperature approaches
the string scale.
This suggests an intriguing connection
with the holographic principle, and leads to some novel speculations
concerning the physics near the critical temperature.

%=============================================================================
\section{Thermal duality and traditional thermodynamics}
\setcounter{footnote}{0}
 
Thermal duality~\cite{OBrienTan,AlvOsoNPB304,AtickWitten,AlvOsoPRD40,OsoIJMP,Polbook}
is a symmetry which relates string thermodynamics at temperature
$T$ to string thermodynamics at the inverse temperature $T_c^2/T$.  Here $T_c$
is a critical, self-dual temperature which is ultimately set by the string scale.
It is easy to see how thermal duality emerges.
In string theory (just as in ordinary quantum field theory), 
finite-temperature effects can be 
incorporated~\cite{Pol86,Polbook}
by compactifying an additional time dimension on a circle 
of radius $R_T = (2\pi T)^{-1}$.
However, Lorentz invariance guarantees that the properties of this
extra time dimension should be the same as those of the original space
dimensions, and T-duality~\cite{SakaiSenda,Nairetal,Sathiapalan}
tells us that closed string theory 
on a compactified space dimension of radius $R$ is indistinguishable
from that on a space of radius $R_c^2/R$ where 
$R_c\equiv \sqrt{\alpha'}= M_{\rm string}^{-1}$ 
is a critical, self-dual radius.  
Together, these two symmetries thus imply a thermal
duality symmetry under which the physics (specifically the one-loop string partition 
function $Z_{\rm string}$) should be invariant with respect to the thermal
duality transformation $T\to T_c^2/T$:  
\beq
           Z_{\rm string} (T_c^2/T) ~=~ Z_{\rm string}(T)~
\label{stringpartfunct}
\eeq
where $T_c\equiv M_{\rm string}/2\pi$.
Note that this symmetry holds to all 
orders in perturbation theory~\cite{AlvOsoPRD40}.

All thermodynamic quantities of interest are generated 
from this partition function.  
The one-loop
vacuum amplitude $\calV(T)$ is given by~\cite{Pol86,McClainRoth,OBrienTan}
\beq
    \calV(T) ~\equiv~ -\half \,{\cal M}^{D-1}\, \int_{\cal F} {d^2 \tau\over ({\rm Im} \,\tau)^2}
             Z_{\rm string}(T)
\label{Vdef}
\eeq
where ${\cal M}\equiv M_{\rm string}/2\pi$ is the reduced string scale;
$D$ is the spacetime dimension;
$\tau$ is the complex modular parameter describing the shape of the
one-loop toroidal worldsheet; 
and ${\cal F}\equiv \lbrace \tau:  |{\rm Re}\,\tau|\leq \half,  
 {\rm Im}\,\tau>0, |\tau|\geq 1\rbrace$ is the fundamental domain
of the modular group.  
In general, $\calV(T)$ plays the role normally assumed by $-\ln Z$ where
$Z$ is the usual thermodynamic partition function in the canonical ensemble.
Given this definition for $\calV$, the free energy $F$, internal 
energy $U$, entropy $S$, and specific heat $c_V$ follow
from the usual thermodynamic definitions:
\beq
          F = T \calV~,~~~~ 
         U = - T^2 {d\over dT} \calV~,~~~~
         S = -{d\over dT} F~,~~~~ c_V = {d\over dT} U~.
\label{usualrelations}
\eeq
It follows directly from these definitions that $U=F+TS$
and that $c_V=T dS/dT$.
Note that $\Lambda \equiv \lim_{T\to 0}F(T)$ is the usual 
one-loop zero-temperature cosmological constant.

It is straightforward to determine the extent to which 
the thermal duality exhibited by $Z_{\rm string}$ in Eq.~(\ref{stringpartfunct})
is inherited 
by its descendants in Eqs.~(\ref{Vdef}) and (\ref{usualrelations}).  
Since $\calV$ is nothing but the modular integral of $Z_{\rm string}$,
the invariance of $Z_{\rm string}$ under the thermal duality 
transformation immediately implies the invariance of $\calV$:
\beq
         \calV(T_c^2/T) ~=~ \calV(T)~.
\eeq
Similarly, from its definition in Eq.~(\ref{usualrelations}),
we find that $F$ transforms {\it covariantly}\/ 
under thermal duality: 
\beq
         F(T_c^2/T) ~=~ (T_c/T)^2\, F(T)~.
\label{Ftrans}
\eeq
Thus, $F$ also respects the thermal duality symmetry.
Finally, it is easy to verify that the internal energy 
$U$ also transforms covariantly under thermal duality:
\beq
         U(T_c^2/T) ~=~ -(T_c/T)^2\, U(T)~.
\label{Utrans}
\eeq
The overall minus sign in this duality transformation
has the net effect of fixing a zero
for the internal energy such that it vanishes at the self-dual
temperature,  $U(T_c)=0$.
Since $dU/dT>0$ for all $T<T_c$,
this zero of energy requires $U(T)<0$ for $T<T_c$.
 
Unfortunately, the entropy 
and specific heat fail to transform either invariantly
or covariantly under the duality transformation.
In other words, these quantities fail to transform
as {\it bona fide}\/ representations of the thermal duality symmetry.
If thermal duality is indeed a fundamental 
property of string theory, the failure of the entropy and
specific heat to transform covariantly under
thermal duality suggests that 
these quantities are improperly defined from a string-theoretic 
standpoint.  At best, they are not the proper ``eigenquantities''
which should correspond to physical observables.

It is straightforward to determine the source of the
difficulty.  Even though $Z_{\rm string}$ and $\calV$ are
thermal duality invariant, the passage to the remaining 
thermodynamic quantities involves the mathematical operations
of multiplication by, and differentiation with respect
to, the temperature $T$.  While multiplication by $T$
preserves covariance under the thermal duality symmetry,
 {\it differentiation with respect to $T$ generally does not}.
Indeed, although the derivative in the definition for $U(T)$
happens to preserve thermal duality covariance,
this covariance is broken by the subsequent
differentiations which are needed to construct the entropy and
specific heat.

The problem of a derivative failing to preserve a symmetry 
is an old one in physics;   {\it the solution is to construct
the analogue of a covariant derivative}.
This procedure is well known in gauge theories,
where the need to construct a covariant derivative
respecting the local gauge symmetry
requires the introduction of an entirely new degree 
of freedom, namely the gauge field.
Fortunately, in the present case of the thermal duality,
the situation is far simpler.

%=============================================================================
\section{Modular invariance and threshold corrections:\\  
       An analogy, and some history}
\setcounter{footnote}{0}

As a digression, let us first consider an analogous
case involving modular invariance.
This case will be mathematically similar to the case of
thermal duality transformations.
In general, a modular-covariant function $f(\tau)$ is
one for which
\beq
         f\left( {a\tau+b\over c\tau+d}\right) ~=~ (c\tau+d)^k f(\tau)
\eeq
for all $a,b,c,d\in\IZ$ with $ad-bc\in\IZ$.
The quantity $k$ is called the modular weight of $f$.
Note that the special case with $(a,b,c,d)=(0,-1,1,0)$ yields
the modular transformation $\tau \to -1/\tau$, which is
very similar to the thermal duality transformation.
However, if $f$ is modular covariant with weight $k$, it
is easy to verify that $df/d\tau$ is not modular covariant;
in other words, $d/d\tau$ is not a modular-covariant
derivative.  Instead, the appropriate modular-covariant
derivative acting on a modular function of weight $k$ is
\beq
     D_\tau ~\equiv ~ {d\over d\tau} ~-~ {ik\over 2\, {\rm Im}\,\tau}~.
\label{modcovderiv}
\eeq
This ensures that if $f$ is a modular-covariant function 
of weight $k$, then $D_\tau f$ is also modular covariant, with
weight $k+2$.
 
The existence of this modular-covariant derivative is not merely 
a mathematical nicety:  it turns out to play an important role
in calculating string threshold corrections to low-energy gauge
couplings~\cite{review}.
Recall that in string theory, the partition function 
$Z_{\rm string}(\tau)$
is a modular-invariant trace over all states in the string Fock space,
where $\tau$ is the complex parameter describing the shape of the torus
(one-loop diagram);
the final result can generally be written
as a sum of products of modular-covariant 
functions $f(\tau)$ and their complex conjugates.  
However, in order to calculate threshold corrections  
to the running of the low-energy gauge couplings due to the infinite
towers of massive string states,
the rules of ordinary quantum field theory
instruct us to calculate a slightly different trace over
the Fock space in which the contribution from each state 
is now multiplied by its squared gauge charge~\cite{Kaplunovsky}.  
However, it turns out that multiplication by the squared gauge charge
in the trace is mathematically equivalent to replacing certain occurrences of $f$
in the final partition function with the derivative $df/d\tau$,
thereby breaking the underlying modular invariance
of the theory.  
Thus, it appears that the usual calculations inherited
from quantum field theory lead to results which fail
to respect the underlying string symmetries.  

This state of affairs persisted for almost
a decade until it was found~\cite{KiritsisKounnas} that
a full {\it string}\/ calculation performed in the presence of a suitable
infrared regulator introduces additional unexpected contributions to the
threshold corrections.   Remarkably, these extra contributions
correspond to adding the second term in Eq.~(\ref{modcovderiv})
to each occurence of $df/d\tau$,
thereby elevating the 
non-covariant derivative $d/d\tau$ 
into the full covariant derivative $D_\tau$.
These extra contributions are intrinsically
gravitational in origin, arising from spacetime curvature 
backreactions and dilaton tadpoles, and thus would not have 
been anticipated from a straightforward field-theoretic derivation. 
However, these extra contributions are precisely what are needed to restore
modular invariance to the full string threshold calculation,
as expected from the string perspective.
A review of this situation can be found in Ref.~\cite{review}.

The lesson from this example is clear:  although field-theoretic considerations
may suggest the existence of certain derivatives in the definitions
of physically relevant quantities, a full string calculation
of these quantities should only involve those covariant forms of these derivatives
which respect the underlying string symmetries.  
What we are proposing, then, is to follow this example
in the case of the temperature derivatives appearing in traditional
string thermodynamics, using thermal duality covariance 
as our guide.  

%=============================================================================
\section{Thermal duality covariant derivatives}
\setcounter{footnote}{0}

We shall now proceed to construct 
our thermal duality covariant derivatives.
We begin with a mathematical definition:
if
a function $f(T)$ has the duality transformation
\beq
           f(T_c^2/T) ~=~ \gamma\, (T_c/T)^k\, f(T)~
\label{covariant}
\eeq
with $\gamma=\pm 1$, we shall say that $f(T)$ is a thermal duality
covariant function with ``weight'' $k$ and sign $\pm 1$ (``even'' or ``odd'').
Note that $\gamma=\pm 1$ are the most general coefficients
which preserve the $\IZ_2$ nature of the thermal duality transformation.

It is easy to verify that multiplication by $T$ is a covariant
operation, resulting in a function with weight $k+2$ and the same
sign for $\gamma$.
Our goal, however, is to construct a thermal duality covariant derivative.
Towards this end, let us imagine that this derivative takes the general form
\beq
     D_T ~=~ {d\over dT} ~+~  {g(T)\over T} 
\label{genform}
\eeq
where $g(T)$ is a function of $T$ and $T_c$.
Explicitly evaluating $[D_T f](T_c^2/T)$ using Eqs.~(\ref{covariant}) and (\ref{genform}),
we then find that 
$D_T f$ will be duality covariant with weight $k-2$ and sign $-\gamma$,
 {\it i.e.}\/,
\beq
                [D_Tf](T_c^2/T) ~=~ -\gamma \left(T_c/T\right)^{k-2} [D_Tf](T)~, 
\eeq
only if $g(T)$ satisfies the constraint
\beq
                 g(T) ~+~ g(T_c^2/T) ~=~ -k~.
\label{constr}
\eeq
Note that $g(T)=0$ is {\it not}\/ a solution if $k\not=0$;  in other words,
for non-zero $k$, we {\it must}\/ make an additional contribution to the ordinary 
temperature derivative in order to preserve duality covariance.
Indeed, since Eq.~(\ref{constr}) must hold for all $T$, 
the function $g(T)$ must be proportional to the weight $k$.
Our task is then to find a suitable function $g(T)$.

In principle, there may be many functions $g(T)$ which satisfy Eq.~(\ref{constr}).
However, again taking duality covariance as our guide, let us 
suppose that $g(T)$ is itself a duality-covariant function with 
weight $\alpha$ and sign $\gamma_g$:
\beq
           g(T_c^2/T) ~=~ \gamma_g\, (T_c/T)^\alpha\, g(T)~.
\label{gcov}
\eeq
Substituting this into Eq.~(\ref{constr}), we then obtain
a solution for $g(T)$:
\beq
           g(T) ~=~ - {k \over 1 + \gamma_g (T_c/T)^\alpha}~=~ 
         - { kT^\alpha\over T^\alpha + \gamma_g T_c^\alpha}~.
\label{gsol}
\eeq
Thus, this solution for $g(T)$ ensures a duality-covariant derivative
for all $\alpha$ and $\gamma_g$.

Thus far, the values of $\alpha$ and $\gamma_g$ are unfixed.
In certain circumstances, however, we can impose 
various physical constraints in order to narrow the range of possibilities.
For example, 
we might wish to demand that our covariant derivative reduce 
to the usual derivative as $T/T_c \to 0$, with only small corrections
suppressed by inverse powers of $T_c$.  
In other words, we wish to demand 
\beq
       { g(T)  f \over T}~\ll~ {df\over dT}~~~~~~~{\rm as}~~T/T_c\to 0~.
\label{smallcorr}
\eeq
With $g(T)$ given by Eq.~(\ref{gsol}), 
this generally restricts us to the cases with $\alpha > 1$,
although this constraint can be evaded or strengthened depending on 
the specific function $f$. 
Likewise, if we wish to retain the usual
symmetry under which the temperature derivative is odd under $T\to -T$,
we should require $\alpha \in 2\IZ$, although once again this 
constraint is not mandatory.
Finally, we would like our covariant derivatives to 
remain finite as
$T\to T_c$.  Thus, we shall restrict our attention to the cases 
with $\gamma_g= +1$, deferring our discussion of the $\gamma_g= -1$ case
to Sect.~7.
We shall, however, leave $\alpha$ as a free (positive) parameter.

Thus, combining our results and taking $\gamma_g= +1$, we obtain
a thermal duality covariant derivative given by
\beq
    D_T ~=~ {d\over dT} ~-~ {k\over T}\, { T^{\alpha}\over T^{\alpha} + T_c^{\alpha} }~.
\label{Ddef}
\eeq
In this derivative, the second term functions as a ``correction'' term which is suppressed
when $T\ll T_c$, but which grows large as the temperature 
approaches the string scale.
Indeed, $\alpha$ essentially governs the {\it rate}\/ at which our correction
term becomes significant as $T\to T_c$.
Of course, the presence of this correction term is critical,
ensuring that if $f$ is covariant 
with weight $k$ and sign $\gamma$, then $D_T f$ is also covariant,
with weight $k-2$ and sign $-\gamma$.
Note that unlike the case with modular transformations, 
there is no thermal duality covariant derivative which
preserves the sign of $\gamma$. 

It may seem strange that our covariant derivative depends on $k$, which
is a property of the function upon which the derivative operates.
However, this is completely analogous to the situation we have just discussed
for modular invariance in Sect.~3.  Indeed, even in gauge theory,
the gauge-covariant derivative depends on the gauge charge of the
state on which it operates.  In this analogy, $k$ functions as the
duality ``charge'' of the function $f(T)$, and the remaining
factor $T^\alpha/(T^\alpha + T_c^\alpha)$ functions as the duality ``gauge field''
(\ie, as the connection).

In principle, the value of $\alpha$ is unconstrained as long as $\alpha > 1$.
We note, however, that in the limit as $T\to T_c$, the covariant derivative
in Eq.~(\ref{Ddef}) takes the limiting form
\beq
          D_T ~\to ~ {d\over dT} ~-~ {k\over 2T}~.
\label{Ddefzero}
\eeq
This is the direct analogue of Eq.~(\ref{modcovderiv}),
and is equivalent to the general derivative in Eq.~(\ref{Ddef})
with $\alpha=0$.
Thus, the $\alpha=0$ case will continue to have relevance at 
the critical temperature $T_c$.
Moreover, as we shall see in Sect.~7, this derivative has another
important property as well.

We stress that this form for the covariant derivative is not unique.
In principle, any function $g(T)$ satisfying Eq.~(\ref{constr}) could serve
in the construction of a covariant derivative.  Of course, physically sensible
solutions for $g(T)$ must have the property that
$g(T)/T\to 0$ as $T/T_c\to 0$, so that our ``corrections'' vanish
at small temperatures and traditional thermodynamics is restored.
Likewise, at the other extreme, we see directly from Eq.~(\ref{constr}) that 
there are only two possibilities as $T\to T_c$:  
either $g(T)$ remains finite, in which case we must have
$g(T)\to -k/2$,  or $g(T)$ diverges, in which case we 
must have $g(T)\to \pm \infty$ as $T\to T_c^\mp$.
In the former case,
we necessarily obtain
the covariant derivative (\ref{Ddefzero}) as $T\to T_c$, regardless of the
specific solution for $g(T)$.  
The specific solution for $g(T)$ therefore serves only to interpolate
between the fixed $T\to 0$ and $T\to T_c$ limits.

Presumably,
the specific form of $g(T)$ [and if $g(T)$ is covariant, the specific value of $\alpha$]
can be determined through a full string 
calculation including gravitational backreactions (analogous to the 
calculation performed in Ref.~\cite{KiritsisKounnas})
in which this covariant derivative is obtained from first principles.  
However, the important point from our analysis is that 
there is {\it necessarily}\/ a string-suppressed ``correction''
term which must be added to the usual temperature derivative,
and that its form is already significantly constrained, especially in 
the $T\to T_c$ limit.
Thus, we shall continue to use the covariant derivative (\ref{Ddef})
in the following, even though we must bear in mind that 
other solutions for $g(T)$ may exist.

%==================================================================================
\section{A duality-covariant string thermodynamics}
\setcounter{footnote}{0}
   
Given this covariant derivative, we can now construct a manifestly 
covariant thermodynamics:  our procedure is simply to replace
all derivatives in Eq.~(\ref{usualrelations})    
with the duality-covariant derivative in Eq.~(\ref{Ddef}).  We thus obtain
\beq
          \tilde F = T \calV~,~~~~ 
         \tilde U = - T^2 D_T \calV~,~~~~
         \tilde S = -D_T \tilde F~,~~~~ \tilde c_V = D_T \tilde U~.
\label{newrelations}
\eeq
The tildes are inserted to emphasize that the new quantities
we are defining need not, {\it a priori}\/, be the same
as their traditional counterparts.

Let us now determine the implications of these definitions.
Of course, since $\calV$ is duality {\it invariant}\/ (\ie,
$\calV$ has $k=0$ with $\gamma= +1$), 
we see that $\tilde F$ continues to be 
covariant with $k=2$ and $\gamma= +1$.
Thus the free energy $F(T)$ is unaltered:  $\tilde F=F$.
This is expected, since we saw in Eq.~(\ref{Ftrans})
that $F$ is already thermal duality covariant.

A similar situation exists for the internal energy $\tilde U(T)$.
Since $\calV$ is covariant with $k=0$, we see that 
the covariant derivative $D_T$ in this special case 
is exactly the same as the usual derivative $d/dT$.  
Thus, the internal energy is also unaffected:  $\tilde U = U$.  
Of course, this also makes sense,
since $U(T)$ was already seen to be covariant 
in Eq.~(\ref{Utrans}), with
$k=2$ and $\gamma = -1$.
However, this example illustrates that any duality-covariant
quantity can (and should) be expressed in terms of covariant
derivatives.  Thus, the internal energy $U$ continues to
fit into our overall framework involving only those derivatives.
  
We now turn our attention to $\tilde S$ and $\tilde c_V$.  
It is in these cases that new features arise.
Since $\tilde F=F $ is already covariant with $k=2$ and $\gamma = +1$,
we find that 
\beq
      \tilde S ~=~ -D_T F ~=~ S ~+~  
              {2\, T^{\alpha-1}\, F \over T^\alpha+ T_c^\alpha}
               ~=~ S~+~ {2\, T^{\alpha} \over T^\alpha+ T_c^\alpha}\, \calV~.
\label{Sdef}
\eeq
Thus, we see that the ``corrected'' entropy $\tilde S$ differs from
the usual entropy $S$ by the addition of an extra string-suppressed term
proportional to the free energy $F$.
Indeed, it is this corrected entropy $\tilde S$ which is 
thermal duality covariant, transforming with $k=0$ and $\gamma = -1$.
Interestingly, since $\tilde S$ is finite and odd, we see that
the corrected entropy has
a zero at the critical temperature:  $\tilde S(T_c)=0$.
This resembles the situation with the internal energy $U$,
which also vanishes at $T=T_c$;
in fact, we find 
\beq
    \tilde S ~\to~ S + {F\over T} = {U\over T}~~~~~~
                {\rm as} ~~ T\to T_c~.
\label{shifts}
\eeq
Of course, both of these properties differ significantly from 
our usual expectations.

Since our corrections to the entropy are suppressed
by powers of the string scale,
we see that $\tilde S$ continues to obey 
the third law of thermodynamics,  with $\tilde S\to 0$ as $T\to 0$ 
in situations with a massless unique ground state.
As discussed in Sect.~4, this is the result of requiring $\alpha>1$.
However, imposing our general condition in Eq.~(\ref{smallcorr}),
we find that we must actually restrict ourselves 
to values of $\alpha$ for which
\beq
       {2\, T^{\alpha-1}\, F \over T^\alpha+ T_c^\alpha} ~\ll ~ S~ 
           ~~~~~~~{\rm as}~~T/T_c\to 0~.
\label{smallcorrS}
\eeq
In general, depending on the particular thermodynamic system under study, this
can yield constraints which are stronger than $\alpha>1$. 

Finally, the corrected specific heat is given by
\beq
      \tilde c_V ~=~ D_T U ~=~ c_V ~-~  {2 \,T^{\alpha-1} \,U\over T^{\alpha} + T_c^{\alpha} }~
      ~=~  c_V ~+~  {2 \,T^{\alpha+1} \over T^{\alpha} + T_c^{\alpha} } \, {d \calV\over dT}~.
\label{cVdef}
\eeq
Thus, the difference between the uncorrected and corrected specific heats 
is a string-suppressed term proportional
to the internal energy $U$. 
Since $U$ vanishes at $T=T_c$ as a result of thermal duality, 
the corrected specific heat $\tilde c_V$
approaches the uncorrected specific heat $c_V$ both 
as $T\to 0$ {\it and}\/ as $T\to T_c$;  indeed, in the latter limit,
we find
\beq
  \tilde c_V ~\to~ c_V - {U\over T} = c_V - \tilde S~
         ~~~~~~~ {\rm as}~~ T\to T_c~. 
\eeq
Note that in general, $\tilde c_V$ is duality invariant and even, 
just like $\calV$.
Once again, for a consistent low-temperature limit which reproduces 
ordinary thermodynamics, we must choose $\alpha$ such that 
\beq
       {2\, T^{\alpha-1}\, U \over T^\alpha+ T_c^\alpha} ~\ll ~ 
           c_V~ ~~~~~~~{\rm as}~~T/T_c\to 0~.
\label{smallcorrcV}
\eeq
This constraint typically yields bounds on $\alpha$ which are the 
same as those stemming from Eq.~(\ref{smallcorrS}).

These new, corrected definitions for entropy and specific heat
restore a certain similarity
between the pairs of thermodynamic quantities 
$(\tilde F,\tilde U)$ and $(\tilde S,\tilde c_V)$.
Members of each pair share the same duality weight  $k$
and have opposite signs for $\gamma$.
Of course, the first pair has
weight $k=2$ while the second pair has $k=0$.

Given these results, we can also see explicitly why  
the usual uncorrected entropy $S$ and specific heat $c_V$ fail to
transform correctly under the thermal duality transformations.  From
Eqs.~(\ref{Sdef}) and (\ref{cVdef}), we see that $S$ and $c_V$
can each be re-expressed  
as admixtures of
$k=0$ quantities which have opposite parities (even or odd)
under the thermal duality transformation.
For example, $S$ is a linear combination of $\tilde S$ (which is
odd) and  
$ T^{\alpha-1} F/(T^{\alpha}+ T_c^{\alpha})$ (a small correction
term which is even).
Only the proper ``corrected'' linear combinations inherent in 
$(\tilde S,\tilde c_V)$ disentangle this behavior.

Thus, we conclude that the ``natural'' duality weights 
and signs for the entropy and specific heat are $k=0$
and $\gamma= \mp 1$ respectively, with the corrections
in Eqs.~(\ref{Sdef}) and (\ref{cVdef})
having the net effect of restoring these properties to an otherwise
non-covariant $S$ and $c_V$.
Moreover, as we have seen,
these transformation properties also make sense from the standpoint
of the usual thermodynamic identities.  
Of course, these conclusions hold only
to the extent that our functions are considered to be
completely general.
For example, as discussed in Ref.~\cite{I},
it is possible to construct special
vacuum amplitudes $\calV(T)$ such that 
the {\it uncorrected}\/ entropy $S(T)$ turns out to
be ``accidentally'' covariant with a non-zero weight.
However, 
even these functions are unsatisfactory because
they are the results of definitions which fail to respect
the thermal duality symmetry.
Thus, such ``accidentally'' covariant entropies
should still be corrected in the manner described here,
thereby restoring
the proper weights and signs to these thermodynamic quantities.
We shall see explicit examples of this below.

In most realistic examples,
the free energy $F$ and the uncorrected entropy $S$ have 
opposite overall signs.
Thus, our string-theoretic corrections to the entropy
in Eq.~(\ref{Sdef})
generally tend to {\it decrease}\/ 
the entropy, 
\beq
               \tilde S ~\leq~ S~,
\eeq
with the suppression becoming increasingly
severe as $T\to T_c$. 
Likewise, in the range $T<T_c$, the internal energy $U$ and the 
specific heat $c_V$ also typically have opposite signs.
[Recall that $U<0$ for $T<T_c$, as discussed below Eq.~(\ref{Utrans}).] 
We therefore find that 
\beq
               \tilde c_V ~\geq~ c_V ~,
\eeq
with the bound saturating both at $T=0$ [where $g(T)=0$] and at $T=T_c$
[where $U(T)=0$].
As we shall see, these inequalities will be extremely important
in the following.

%=============================================================================
\section{An explicit duality-covariant example}
\setcounter{footnote}{0}

In this section, we shall calculate the string-corrected entropy
and specific heat 
within the context of a specific example displaying
thermal duality.
Towards this end, let us consider
the vacuum amplitude~\cite{I} 
\beq
      \calV^{(D)}(T) ~=~  - {(T^D+T_c^D)^{2/D}\over T T_c}
\label{Vell}
\eeq
where $D\geq 1$ is an arbitrary exponent.
(Here and henceforth, we shall express all thermodynamic quantities
in units of the reduced string scale ${\cal M}\equiv M_{\rm string}/2\pi$.) 
As discussed in Ref.~\cite{I}, this function $\calV(T)$ 
emerges as the vacuum amplitude  corresponding
(either exactly, or approximately and highly accurately) 
to finite-temperature string constructions
in which the time/temperature dimension
is compactified on a circle.  The parameter $D$
is the spacetime dimension of the zero-temperature string model.
Note that this functional form for $\calV$ has the property that
the resulting entropy $S$ is ``accidentally'' covariant with 
weight $D$ and sign $+1$;  however, this will play no role in the following.
Indeed, the corresponding specific heat is non-covariant for all $D>2$.

\subsection{String-corrected entropy}

Given this functional form for $\calV(T)$, it is straightforward to calculate 
both the traditional entropy $S(T)$ 
and the corrected entropy $\tilde S(T)$ as functions of temperature, obtaining
\beq
   S^{(D)}(T) ~=~  2 \,{T^{D -1}\over T_c}\, (T^D + T_c^D)^{2/D -1}~
\label{Scov}
\eeq
and 
\beq
   \tilde S^{(D)}(T) ~=~ 2 \,{(T^D + T_c^D)^{2/D-1}
                    \over
                   T T_c \,(T^\alpha + T_c^\alpha)}\,
                 (T^D T_c^\alpha - T^\alpha T_c^D)~.
\label{tildeScov}
\eeq
Note, in particular, that the relative sizes of the string corrections 
to the entropy are not small in this example unless $\alpha \geq D+1$.  
This is the strengthened bound on  $\alpha$ which emerges 
from Eq.~(\ref{smallcorrS}) for this system.

%================== FIGURE ============================================
\begin{figure}[ht]
\centerline{
   \epsfxsize 3.2 truein \epsfbox {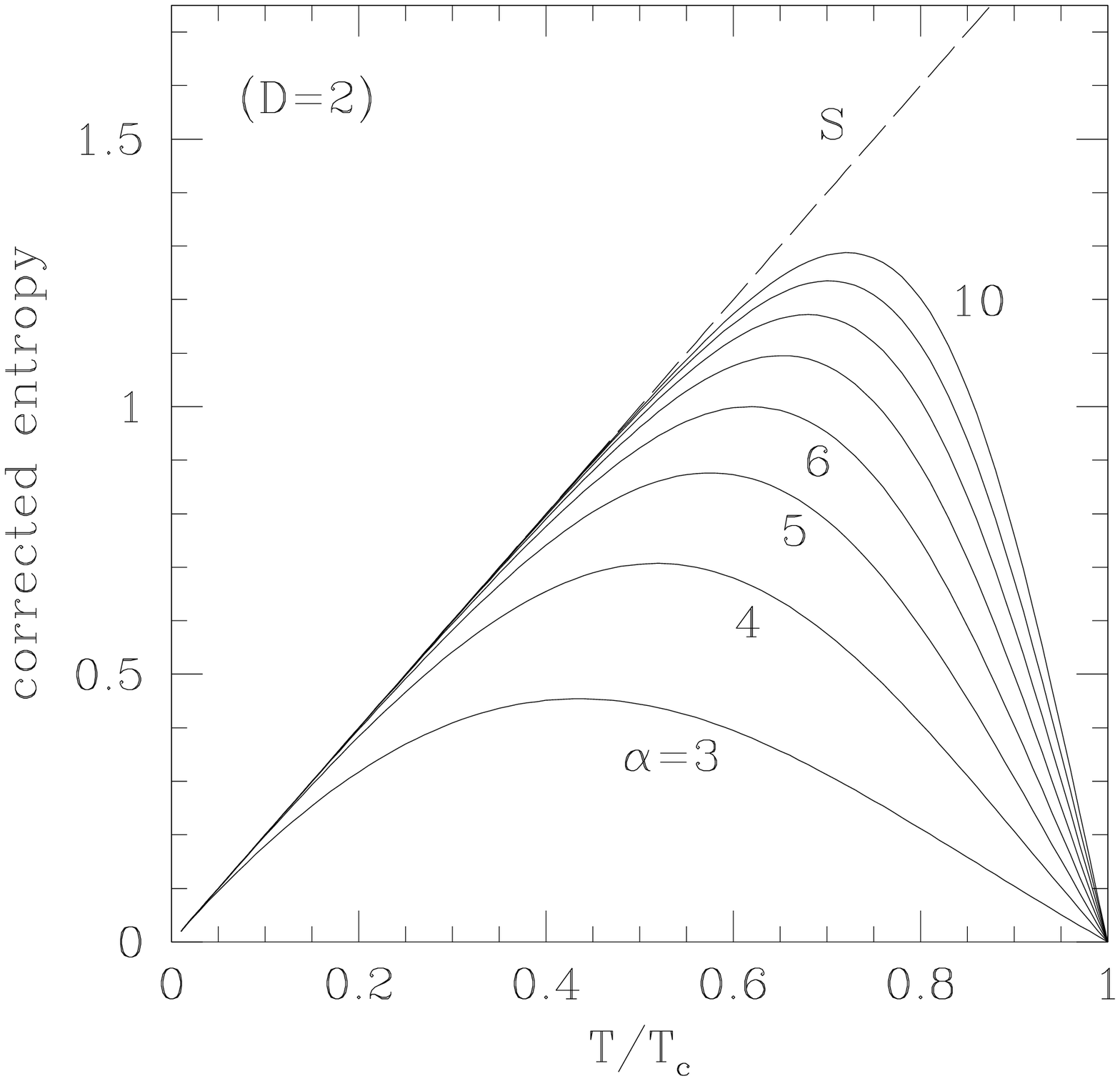}
   \epsfxsize 3.2  truein \epsfbox {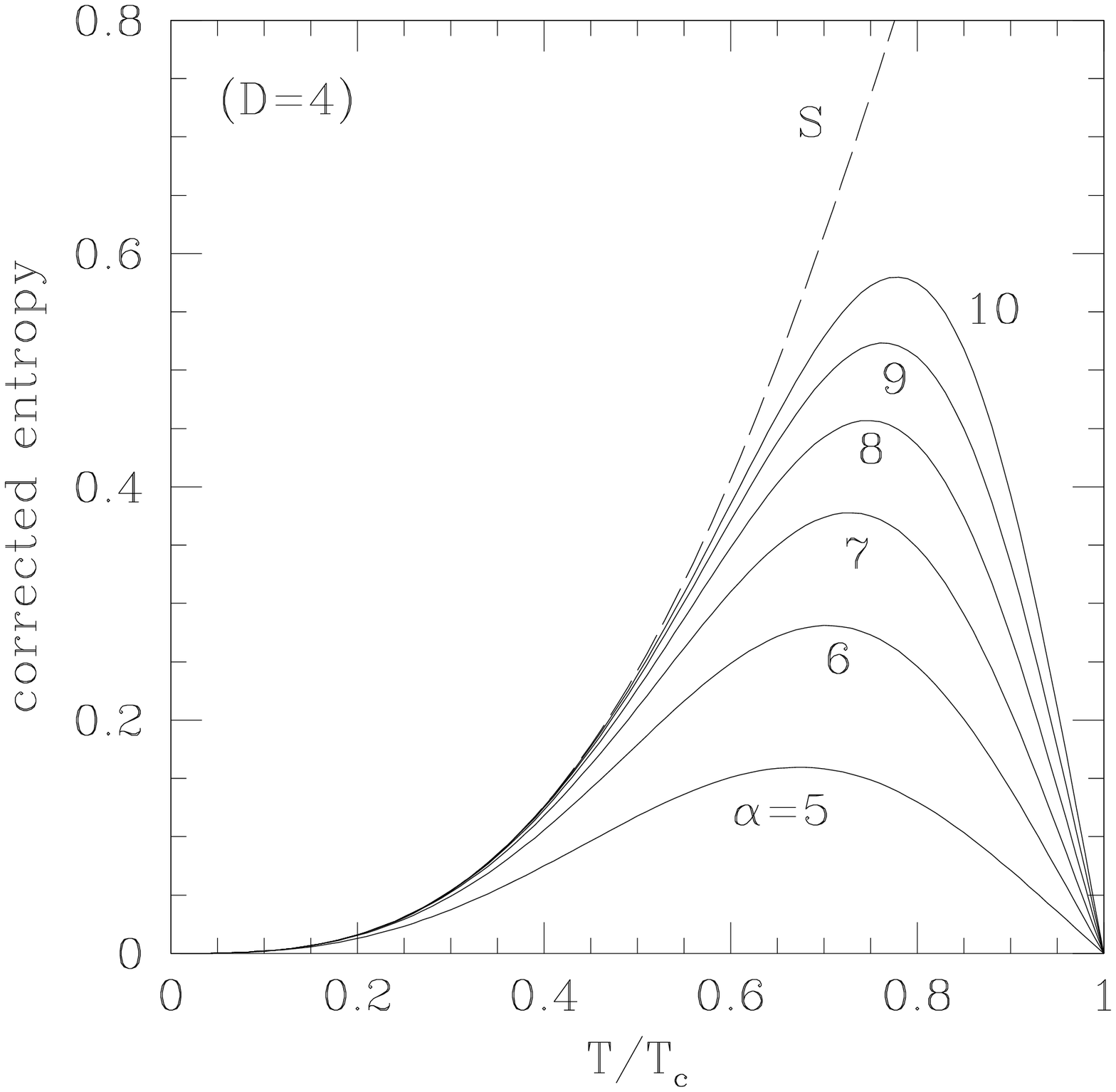}
    }
\caption{The string-corrected entropies $\tilde S^{(D)}$
   given in Eq.~(\protect\ref{tildeScov}),
   plotted as functions of $T$ for $D=2$ (left plot) and $D=4$ (right plot).
   In each case, we have plotted the string-corrected entropies for $D+1\leq \alpha \leq 10$,
    while the uncorrected entropy is indicated with a dashed line.
    In all cases, the corrected entropies are smaller than the traditional
    entropies, and vanish as $T\to T_c$.  The parameter $\alpha$ governs the relative
    size of the string corrections and thus the rate
    with which the corrected entropy begins to separate from the uncorrected entropy.}
\label{fig1}
\end{figure}
%================== END OF INSERTED FIGURE ============================

These functions for $S^{(D)}(T)$ and $\tilde S^{(D)}(T)$ 
are plotted in Fig.~\ref{fig1}
for the special cases with $D=2$ and $D=4$. 
We can immediately see the behavior
of $\tilde S^{(D)}(T)$ as a function of $T$.
At low temperatures $T\ll T_c$, we see that $\tilde S$ follows $S$
quite closely;  indeed the ``corrections'' to the traditional
entropy [\eg, as measured by the ratio $(S-\tilde S)/S$] 
remain small as long as $T\ll T_c$.
At higher temperatures, however,
$\tilde S$ is increasingly
suppressed relative to $S$, and ultimately vanishes as $T\to T_c$.
This is required by the fact that $\tilde S$ must be an odd function 
under $T\to T_c^2/T$ for all $\alpha$.

For sufficiently small temperatures, our corrected entropies 
resemble the traditional entropy and 
grow with increasing temperature, with
$d\tilde S^{(D)}/dT > 0$
all $D$ and $\alpha$.
This conforms to our standard notions of entropy as 
a measure of disorder.
However, as $T$ approaches the critical temperature, we see that
$d\tilde S^{(D)}/dT$ ultimately changes sign.  
At first glance, this might appear to signal an inconsistency
in our string-corrected thermodynamics.
However, as is well known in string thermodynamics (see, \eg,
Refs.~\cite{Hagedorn,Huang,Bowick,Tye,Alvarez,Sathiapalan2,AlvOsoPRD36,AtickWitten,LeBlanc,Deo,BowickGiddings,Giddings,AntonKounnas}),
we expect that a phase transition or other Hagedorn-related event 
should occur at large temperatures at or near $T_c$.
Thus, rather than interpret
$d\tilde S^{(D)}/dT < 0$
as a loss of disorder, it is tempting to interpret this sign change
as the beginning of a possible phase transition and the 
conversion of the system into new degrees of freedom.
Thus, as the temperature increases towards the critical temperature,
fewer and fewer of the original degrees of freedom 
remain in the system, and thus the entropy associated with these
original degrees of freedom begins to decrease.

Of course, verifying this speculation would require a more complete understanding
of the nature of the string physics near the critical temperature. 
Our point here, however, is that a fully covariant treatment of
entropy necessarily requires the introduction of corrections which,
in this case, ultimately drive the corrected entropy to zero at the 
critical temperature.

\subsection{String-corrected specific heat}

\begin{figure}[t]
\centerline{
   \epsfxsize 3.2 truein \epsfbox {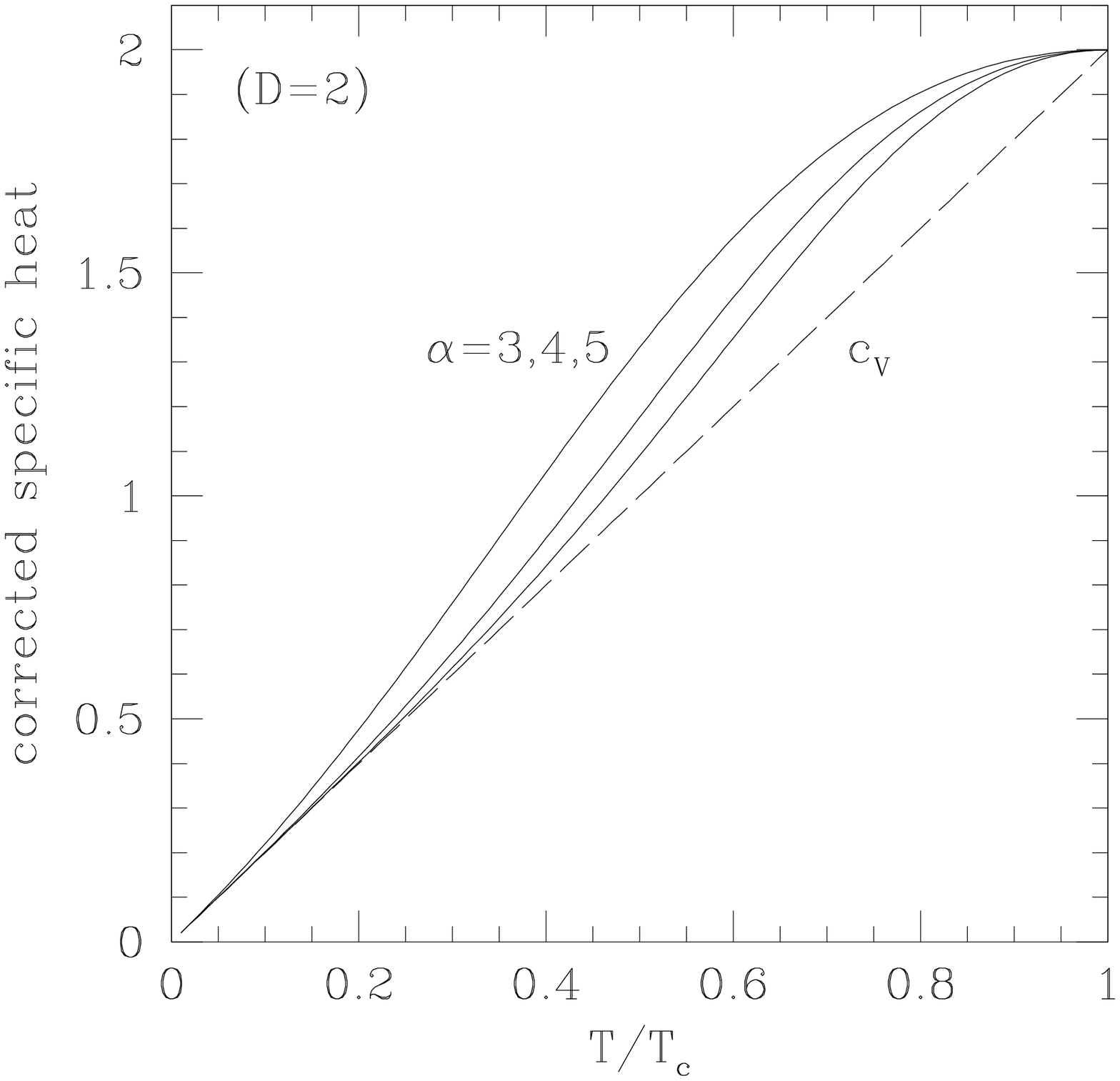}
   \epsfxsize 3.2  truein \epsfbox {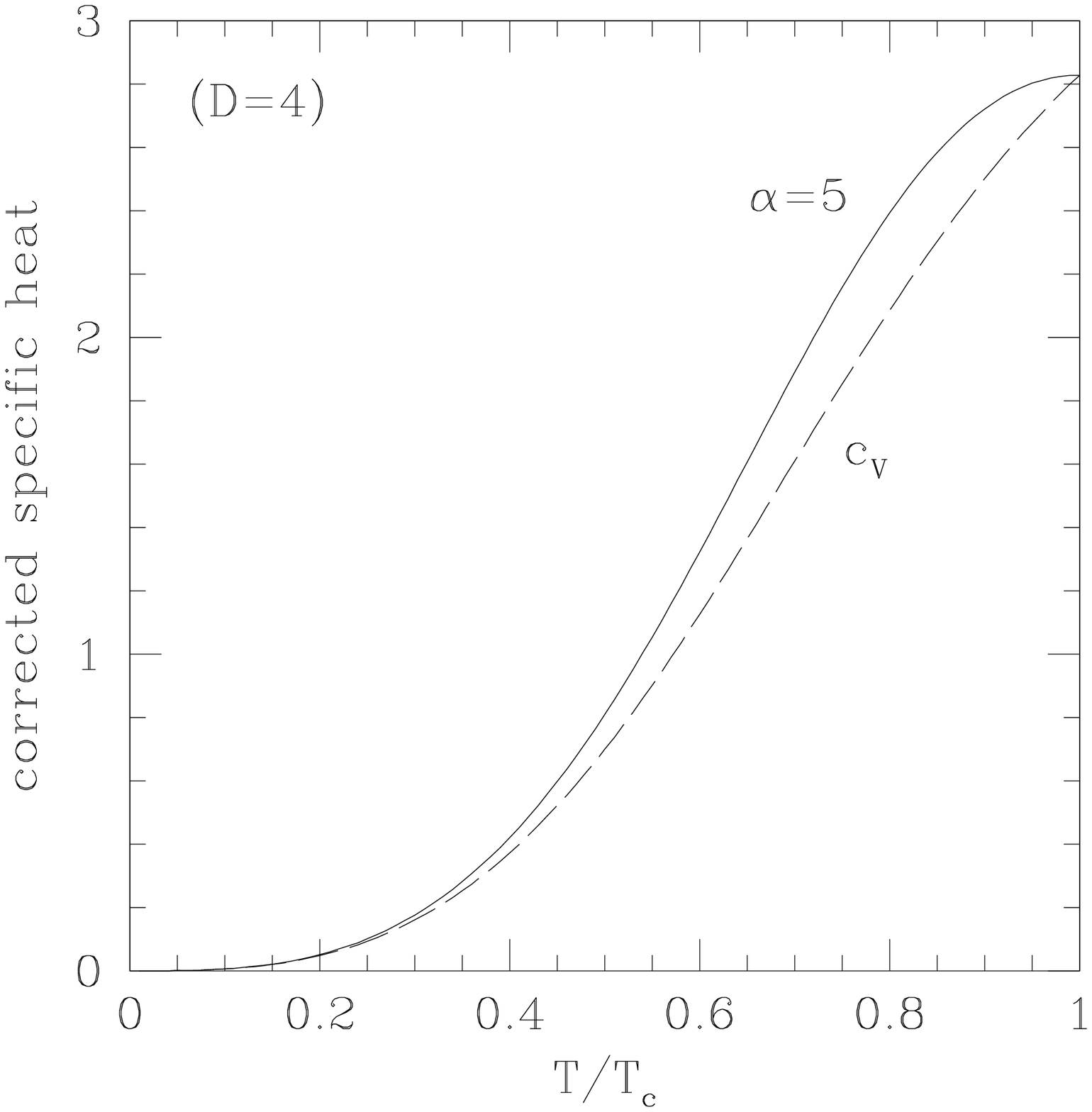}
    }
\caption{The string-corrected specific heats $\tilde c_V^{(D)}$
   given in Eq.~(\protect\ref{tildecVcov}),
   plotted as functions of $T$ for $D=2$ (left plot) and $D=4$ (right plot).
   For the sake of clarity, we have illustrated only the cases with the largest
    relative corrections:  we have taken $\alpha=3,4,5$ for $D=2$, 
    and $\alpha=5$ for $D=4$.
    Note that in all cases, the corrected specific heats exceed the traditional
    specific heat, agreeing with the traditional heat only at $T=0$ and $T=T_c$.
    Moreover, as a consequence of thermal duality, we 
    have $d \tilde c_V^{(D)}/dT=0$ at $T=T_c$ for all $\alpha$.  }
\label{fig2}
\end{figure}

We can also perform a similar analysis for the specific heat.
Once again starting from Eq.~(\ref{Vell}), we obtain
\beq
        c^{(D)}_V(T) ~=~ 2 \,{T^{D-1}\over T_c} \,(T^D + T_c^D)^{2/D -2}
         \left\lbrack   T^D + (D -1)\, T_c^D \right\rbrack
\label{cVcov}
\eeq
and
\beqn
        \tilde c^{(D)}_V(T) &=& 2\, 
      { (T^D+T_c^D)^{2/D-2}\over   T T_c \,(T^\alpha+T_c^\alpha)}\,
          \biggl\lbrack
          T^{2D} T_c^\alpha + T^\alpha T_c^{2D} \nonumber\\
        & & ~~~~~~~~~~~~~~~~~~~~~~~~~~~+(D-1)\, (T^D T_c^{\alpha+D} +  
              T^{\alpha+D} T_c^D)\biggr\rbrack~.
\label{tildecVcov}
\eeqn
These functions are plotted in Fig.~\ref{fig2}
for the special cases with $D=2$ and $D=4$. 
Once again, we observe that $\tilde c_V^{(D)}\geq c_V^{(D)}$, with
the bound saturating only at $T=0$ and $T=T_c$. 
Note that $d \tilde c_V^{(D)}/dT=0$ at $T=T_c$ for all $\alpha$.  
This is a direct consequence of thermal duality, and indicates
that the corrected specific heat loses all temperature sensitivity
at $T_c$.

%================== FIGURE ============================================
%================== END OF INSERTED FIGURE ============================

We conclude this section with two further comments. 
First, throughout this section, we have focused exclusively on the behavior
for $T\leq T_c$.  As mentioned above, we have done this in the expectation 
that a phase transition or other Hagedorn-related event should occur at large 
temperatures at or near $T_c$.  However, from a purely mathematical perspective, 
we could easily have continued our analysis beyond $T_c$, since our string-corrected 
entropies and specific heats are (by construction) thermal duality invariant.  
For example, since the specific heat is necessarily 
an even (invariant) function under $T\to T_c^2/T$, 
we see that $\tilde c_V^{(D)}$ continues to remain positive 
for all $T$ and ultimately declines beyond $T_c$.
This is in sharp contrast to the uncorrected specific heat, which continues to rise
indefinitely.
On the other hand, the string-corrected entropy $\tilde S^{(D)}$ 
is necessarily an odd function under $T\to T_c^2/T$.
Thus, $\tilde S^{(D)}$ becomes {\it negative}\/ beyond $T_c$.
This provides dramatic illustration of the fact that, as already anticipated from
other considerations, new physics must intercede at or near the string scale.

Our second comment concerns the duality weights of the entropy and 
specific heat.  As already mentioned at the beginning of this section
(and as explained more fully in Ref.~\cite{I}), the {\it uncorrected}\/ 
entropy $S^{(D)}$ in this example is actually already covariant,
with duality weight $D$ and sign $+1$.
Thus, it may seem that no further corrections are necessary in this case.
However, as we have seen in Sect.~5, the proper duality weight and sign
for the entropy are $k=0$ and $\gamma= -1$ respectively. 
Thus, the net effect of our corrections in this case is to 
``convert'' an even entropy function of weight $D$ into an 
odd entropy function of weight zero.
Of course, these corrections also simultaneously restore
duality invariance to the specific heat, where it was otherwise lacking.

%=============================================================================
\subsection{Effective dimensionalities and holography}
\setcounter{footnote}{0}

Finally, we now investigate the {\it scaling}\/ behavior 
of our corrected thermodynamic 
quantities as functions of 
temperature.  As we shall see, this will enable us to provide a possible 
physical interpretation to our string-theoretic corrections.

In ordinary quantum field theory, the free energy $F(T)$ at large
temperatures typically scales like $T^{D}$ where $D$ is the spacetime
dimension.  This in turn implies that the entropy $S$ should scale like $T^{D-1}$.
However, in string theory we have $F(T)\sim T^2$ as $T\to \infty$,
implying that $S(T)\sim T$ as $T\to \infty$.
Thus, string theory behaves asymptotically as though it has
an effective dimensionality $D_{\rm eff}=2$.

At first glance, these two sets of results might not appear to be
in conflict since they apply to different theories.
However, the field-theory limit of string theory is expected
to occur for $T\ll T_c$, and thus the field-theory behavior must
be embedded within the larger string-theory behavior.
It is therefore interesting to examine the effective dimensionality
(\ie, the effective scaling exponent)
of our thermodynamic quantities as a function of temperature.
As discussed in Ref.~\cite{I},
it is easiest to define this
effective dimensionality $D_{\rm eff}(T)$ by considering the entropy:
since $S(T)$ is a monotonically increasing function of $T$,
we can define $D_{\rm eff}(T)$ as the effective scaling exponent
at temperature $T$, setting
$S(T)\sim   T^{D_{\rm eff}-1}$.
We thus obtain 
\beq
    D_{\rm eff} ~\equiv~  1+ { d \ln S\over d\ln T} ~=~
               1 +  {T\over S} {dS\over dT} ~=~ 1 + {c_V\over S}~,
\label{Deff}
\eeq 
where the last equality follows from the thermodynamic
identity $c_V= T dS/dT$.

Given the entropy $S^{(D)}$ in Eq.~(\ref{Scov}) and 
the specific heat $c_V^{(D)}$ in Eq.~(\ref{cVcov}), 
it is straightforward to calculate $D_{\rm eff}(T)$ as a 
function of temperature $T$.
This calculation was originally performed in Ref.~\cite{I}, where a plot
of $D_{\rm eff}(T)$ is given. 
In each case, it is found that $D_{\rm eff}$ {\it interpolates}\/ between
$D_{\rm eff}=D$ for $T\ll T_c$
and $D_{\rm eff}=2$ for $T\gg T_c$.
It is, of course, easy to interpret this result.
At small temperatures $T\ll T_c$, the entropy
behaves as we expect on the basis of field theory, growing according
to the power-law $S^{(D)}(T)\sim T^{D-1}$.
Indeed, this low-temperature limit of string theory can be 
identified as the high-temperature limit of the low-energy 
effective field theory.
However, as $T$ approaches the
string scale $T_c$,
we see that this scaling behavior begins to change,
with the $T^{D-1}$ growth in the entropy ultimately becoming
the expected {\it linear} growth for $T\gg T_c$.
This is then the  asymptotic {\it string}\/ limit.

These observations originally appeared in Ref.~\cite{I}.
However, given these observations, let us now proceed to determine 
the effective dimensionalities $\tilde D_{\rm eff}$
of our {\it string-corrected}\/ entropies.  In complete analogy 
with $D_{\rm eff}$, these corrected effective dimensionalities
$\tilde D_{\rm eff}$ may be defined as 
\beq
    \tilde D_{\rm eff} ~\equiv~  1+ { d \ln \tilde S\over d\ln T} ~=~
               1 +  {T\over \tilde S} {d\tilde S\over dT} ~.
\label{tDeff}
\eeq
Note that since $\tilde c_V\not= T d\tilde S/dT$, 
we cannot write Eq.~(\ref{tDeff}) easily in terms of $\tilde c_V$.

%================== FIGURE ============================================
\begin{figure}[ht]
\centerline{
   \epsfxsize 3.0 truein \epsfbox {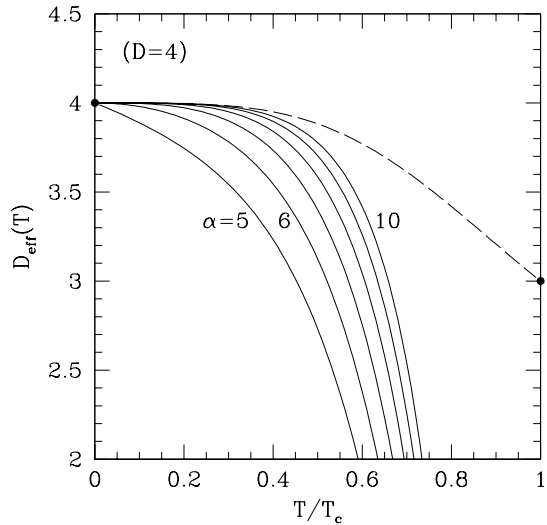}
    }
\caption{The effective dimensionalities $\tilde D_{\rm eff}$
    of the four-dimensional string-corrected entropies $\tilde S^{(4)}$,
    plotted as functions of $T$
   for $5 \leq \alpha \leq 10$. 
   The effective dimensionality of the uncorrected entropy $S^{(4)}$ is
    also shown (dashed line).} 
\label{fig3}
\end{figure}
%================== END OF INSERTED FIGURE ============================

The results are plotted in Fig.~\ref{fig3} for $D=4$.
As expected, all of our corrected entropies exhibit an initial scaling
with $\tilde D_{\rm eff}=D=4$ as $T/T_c\to 0$;  
this is guaranteed  by our original requirement that $\alpha \geq D+1$.
This implies that none of our string corrections disturb
the expected field-theoretic behavior at low temperatures. 
However, as $T$ becomes larger and approaches the string scale, we 
see that the net effect of
our string corrections is {\it to reduce the effective scaling dimensionality 
of the entropy even more rapidly than in the uncorrected case}\/.

Thus, combining our results from Figs.~\ref{fig1} and \ref{fig3}, we see
that 
our string corrections have two net effects on the entropy as $T\to T_c$:
they reduce its overall magnitude, and they also reduce its scaling exponent
(effective dimensionality) as a function of temperature.  
It is important to stress that these are, in principle, uncorrelated
effects:  the first relates to the overall size of a function,
while the second has to do with its rate of growth.
As a stark example of this point, observe that if the scaling behavior of
the corrected entropy had been $(T/T_c)^{3}$ 
rather than $(T/T_c)^4$ for all $T\leq T_c$, this {\it decrease}\/ in the scaling exponent 
would have resulted in an {\it increase}\/ in the entropy, not a decrease. 
This would have been interpreted as the appearance of more degrees of freedom
at low temperatures, not fewer.

Of course, there is a natural interpretation for an effect which simultaneously
decreases not only the entropy but also the effective dimensionality that 
governs its scaling:  such an effect is holographic.
Thus, we see that our duality-inspired corrections to the laws of thermodynamics
are holographic in nature, enhancing the tendency towards
holography that already exists in traditional string thermodynamics.
Indeed, as originally observed in Ref.~\cite{I},
we see from Fig.~\ref{fig3} 
that the {\it uncorrected}\/ effective dimensionality already shows
a holographic decline from $D_{\rm eff}=4$ at $T/T_c\ll 1$ to 
$D_{\rm eff}=3$ at $T\to T_c$.
Our corrections thus enhance this effect, introducing this holographic
behavior even more strikingly and at lower temperatures.

Of course, as discussed more fully in Ref.~\cite{I}, there are a number 
of outstanding issues that need to
be addressed before we can truly identify this phenomenon with holography.
In particular,  an analysis formulated
in flat space (such as ours) cannot address
questions pertaining to the geometry of holography,
and thus cannot determine whether the modified scaling behavior 
and the implied reduction in the number of associated degrees
of freedom are really to be associated with a lower-dimensional 
subspace (or boundary) of the original geometry.
Indeed, such an analysis is beyond the scope of this paper,
and would require reformulating the predictions of thermal
duality for string theories in non-trivial $D$-dimensional backgrounds,
and then developing a map between
degrees of freedom in the bulk of the $D$-dimensional volume
and those on a lower-dimensional section of this volume. 
Thus, as indicated in Ref.~\cite{I}, the possible
connection between thermal duality and holography  
remains to be explored further.

%=============================================================================

\section{Alternative formulations for a duality-covariant\\ thermodynamics}
\setcounter{footnote}{0}

In this section, we shall investigate other possible formulations for
a duality-covariant thermodynamics.  As we shall see, a wide set of 
possibilities exists:  some of these lead to drastically different 
phenomenologies, while others have drastically different theoretical 
underpinnings.

\subsection{Alternative covariant derivatives}

First, as stressed in Sect.~4, our thermal duality covariant derivative
in Eq.~(\ref{Ddef}) is not unique:  {\it any}\/ function $g(T)$ satisfying
Eq.~(\ref{constr}) can be exploited in the construction of a covariant
derivative as in Eq.~(\ref{genform}).  
As an example, let us again remain within the class of covariant functions $g(T)$
given in Eq.~(\ref{gsol}) 
and consider the physics that results if we choose $\gamma_g= -1$
rather than $\gamma_g= +1$.
Our covariant derivative in Eq.~(\ref{Ddef}) then becomes 
\beq
    D^{(-)}_T ~=~ {d\over dT} ~+~ {k\over T}\, { T^{\alpha}\over T_c^{\alpha} - T^{\alpha} }~,
\label{Ddefminus}
\eeq
leading to the definitions
\beqn
     \tilde S &\equiv&  -D^{(-)}_T F ~=~ S ~-~
             {2\, T^{\alpha-1}\, F \over T_c^\alpha- T^\alpha}
              ~=~
              S~-~
             {2\, T^{\alpha} \over T_c^\alpha- T^\alpha}\, \calV~,\nonumber\\
    \tilde c_V &\equiv & D^{(-)}_T U ~=~ c_V ~+~  {2 \,T^{\alpha-1} \,U\over T_c^{\alpha} - T^{\alpha} }~
      ~=~  c_V ~-~  {2 \,T^{\alpha+1} \over T_c^{\alpha} - T^{\alpha} } \, {d \calV\over dT}~.
\label{otherScV}
\eeqn
Note that unlike the case with $\gamma_g=+1$, we now have $\tilde S\geq S$ and 
$\tilde c_V\leq c_V$. 
However, as required, we still find that our string-corrected quantities
$\tilde S$ and $\tilde c_V$ are duality covariant 
with weight $k=0$ and signs $\mp 1$ respectively.
Moreover, in the case of the covariant example given in Eq.~(\ref{Vell}),
a proper low-temperature (field-theory) limit is guaranteed for all $\alpha\geq D$. 

At first glance, the definitions in Eq.~(\ref{otherScV}) 
might appear to be unacceptable because of the
apparent divergences in $\tilde S$ and $\tilde c_V$ as $T\to T_c$.
For example, since the corrections to the entropy are positive
and the corrections to the specific heat are negative,
we might worry that 
the definitions in Eq.~(\ref{otherScV}) would result in the
asymptotic behavior $\tilde S\to \infty$
and $\tilde c_V\to -\infty$ as $T\to T_c$.
While a positively divergent entropy leads to no specific difficulty
(and might be interpreted as a Hagedorn-like phenomenon),
a negative specific heat necessarily results in an inconsistent thermodynamics
in which thermal fluctuations grow without bound and 
ultimately destabilize the system.

However, these concerns are ultimately spurious.
Because the internal energy $U$ vanishes at $T=T_c$ as a result
of thermal duality,
the specific heat actually remains finite and positive as $T\to T_c$.
Indeed, the divergence in the definition of the
covariant derivative cancels against the
vanishing of the internal energy,
resulting in a string-corrected specific 
heat which takes the finite asymptotic value
\beq
       \tilde c_V^{(D)} ~\to~     2^{2/D -1} \,D \,(1-2/\alpha)
       ~~~~~~~ {\rm as}~~ T\to T_c~.
\eeq
Note that this quantity is positive for all $\alpha >2$.

%================== FIGURE ============================================
\begin{figure}[ht]
\centerline{
   \epsfxsize 3.2 truein \epsfbox {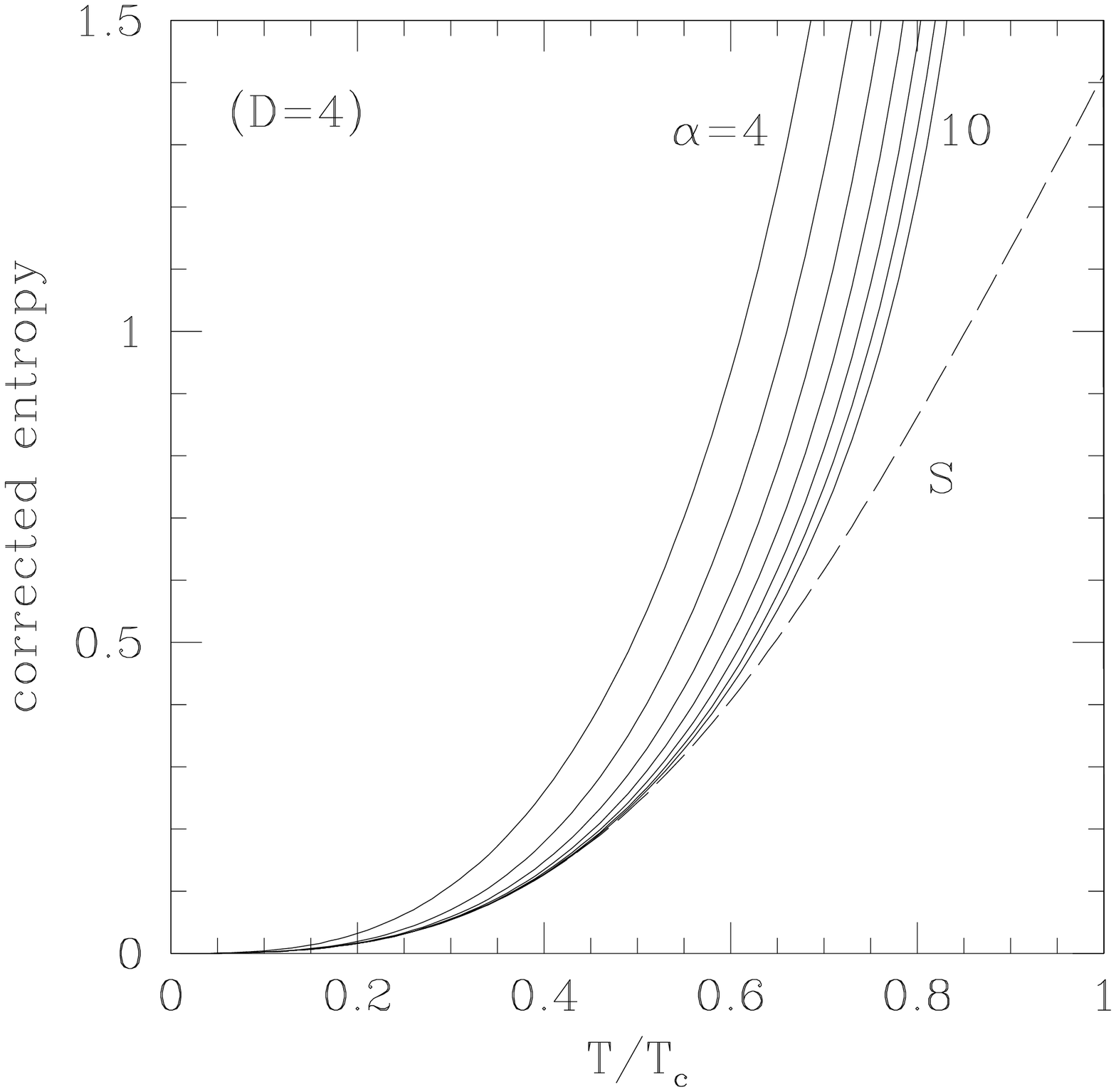}
   \epsfxsize 3.2  truein \epsfbox {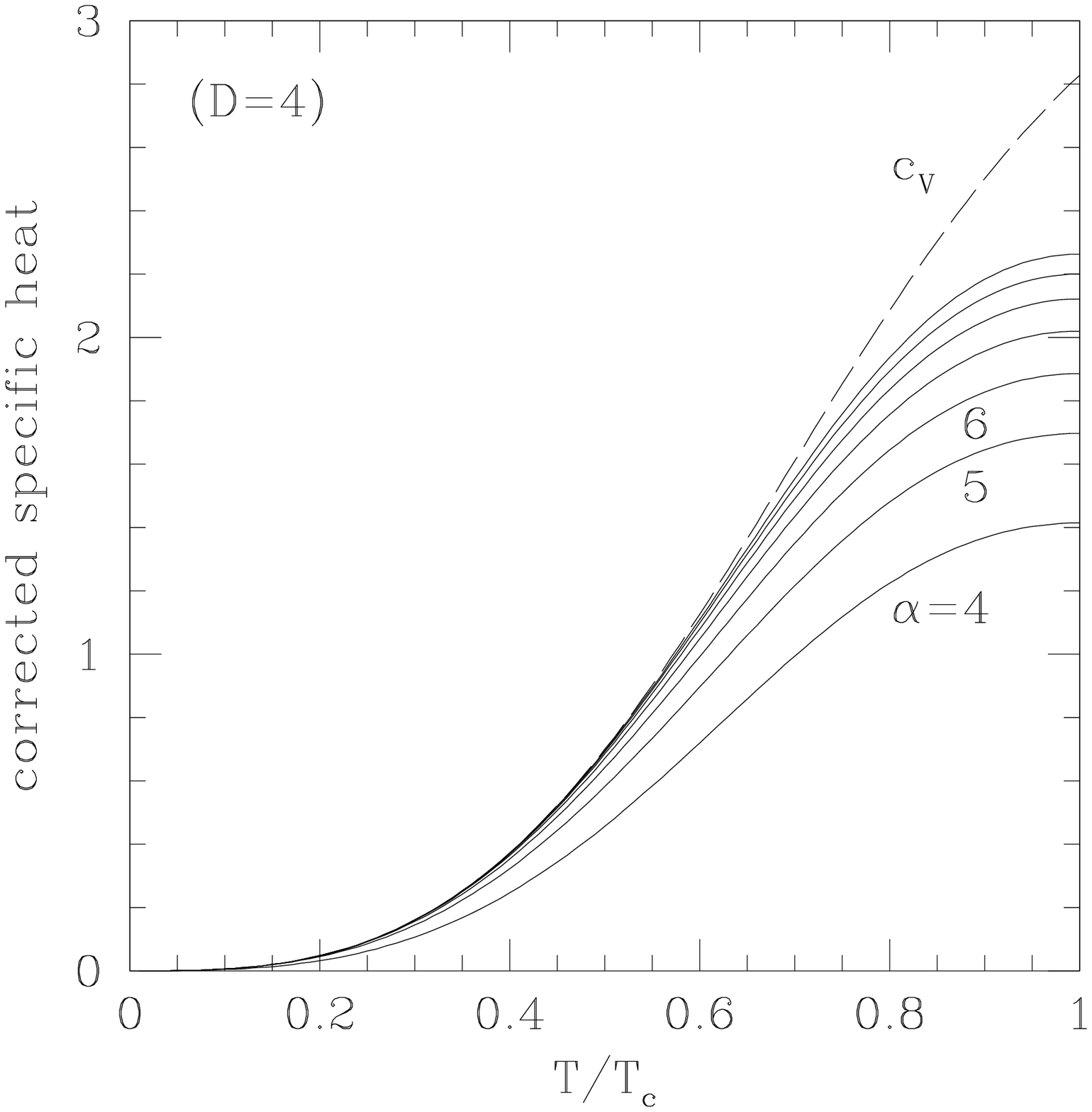}
    }
\caption{The string-corrected four-dimensional entropies 
         $\tilde S^{(4)}$ and specific heats
     $\tilde c_V^{(4)}$ 
   given in Eq.~(\protect\ref{otherScV}),
   plotted as functions of $T$.
   In each case, we have plotted these thermodynamic quantities 
     for $4\leq \alpha \leq 10$,
    while the uncorrected quantities are indicated with a dashed line.
    Note that $\tilde c_V^{(D)}$ remains positive in all cases,
    while $\tilde S^{(D)}$ is now monotonically increasing as a function
    of temperature for all $T\leq T_c$.}
\label{fig4}
\end{figure}
%================== END OF INSERTED FIGURE ============================

The resulting string-corrected entropies and specific heats are plotted in 
Fig.~\ref{fig4} for $D=4$.
As expected, we see that $\tilde c_V^{(4)}$ remains positive in 
all cases, while $\tilde S^{(4)}$ is now monotonically increasing 
as a function of temperature for all $T\leq T_c$.
Clearly, the effect of these corrections is no longer ``holographic''
as it was for $\gamma_g= 1$.  However, this possibility also
results in a fully consistent, duality-covariant string thermodynamics.

At present, we have no physical basis on which to prefer one version of 
the covariant thermodynamics over another.  
Even though they lead to drastically different phenomenologies,
they are each internally self-consistent and have the same 
low-temperature (field-theoretic) limits.
However, our main point in this paper is that {\it some}\/
string-theoretic correction is necessary in order to restore
thermal duality covariance to the usual rules of thermodynamics,
and that it is possible to introduce such corrections
without disturbing the usual
low-temperature physics associated with traditional thermodynamics.
The decision as to the preferred specific form of the covariant derivative
awaits a full string calculation, perhaps along
the lines discussed in Sect.~3.

\subsection{Alternative thermodynamic structures}

Changing the specific form of the covariant derivative 
is not the only way in which we might approach the construction
of an alternative thermodynamics.  Indeed, even within the context of a
fixed covariant derivative, there are other structural 
options that can be explored.

\begin{figure}[t]
\centerline{
   \epsfxsize 2.6 truein \epsfbox {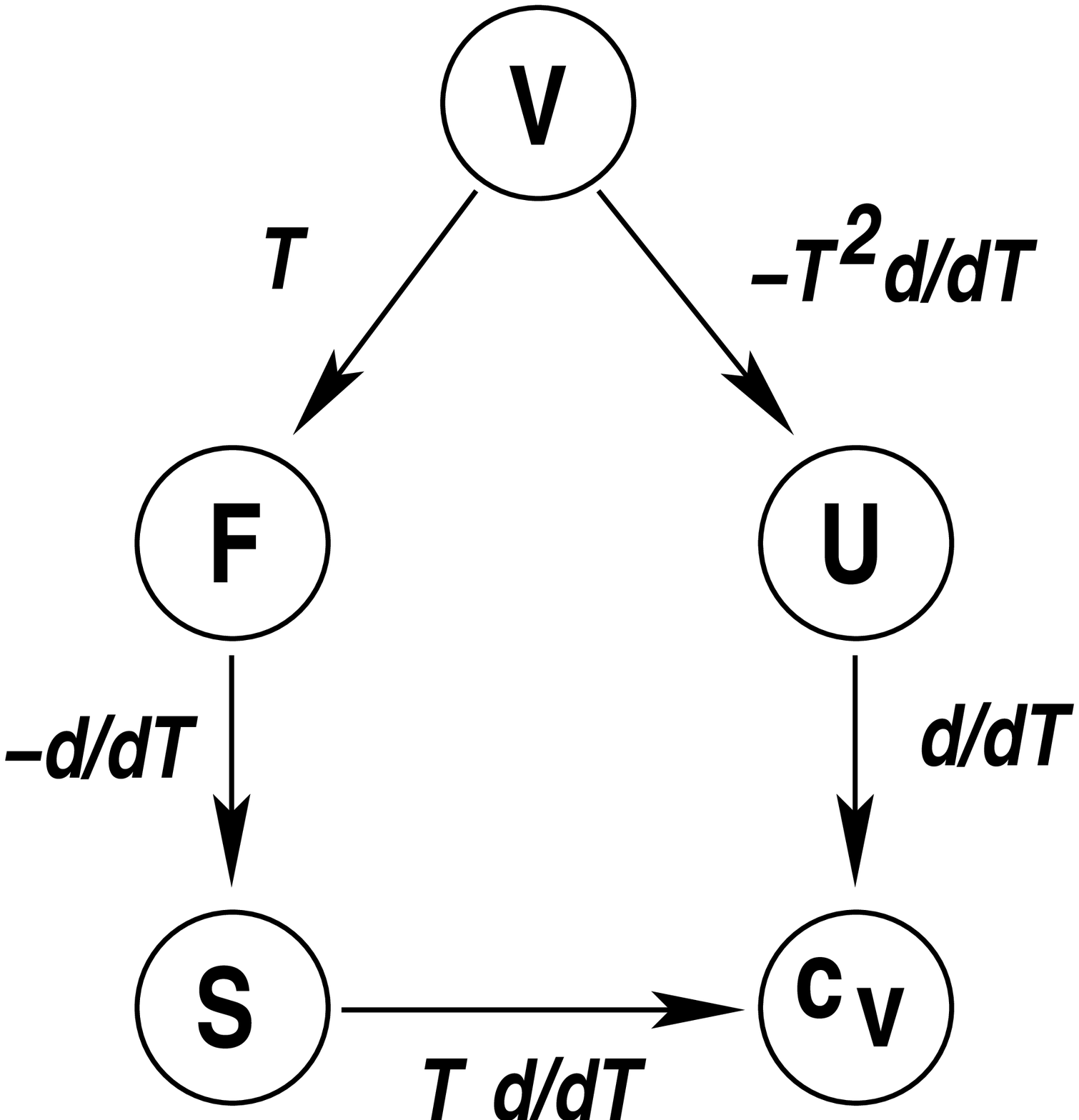}
   \hfill
   \epsfxsize 2.6  truein \epsfbox {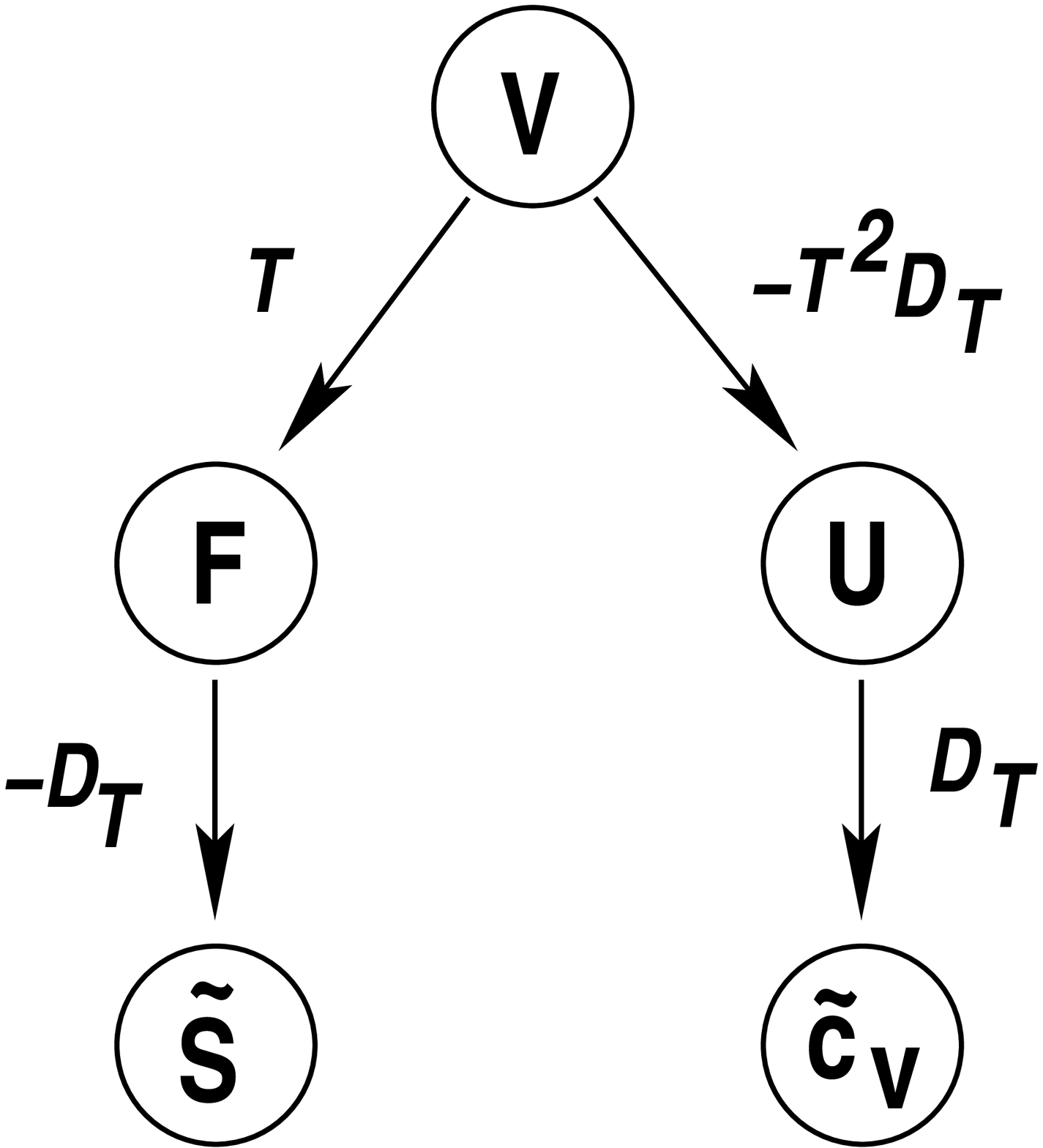}
    }
\caption{ 
   Relations between thermodynamic quantities. 
    (a) Traditional thermodynamics:
        All thermodynamic quantities are related to each other through
        temperature multiplications and differentiations.
    (b) Our string-corrected thermodynamics:  we replace the usual 
        temperature derivatives by duality-covariant derivatives,
        maintaining the definitions of $\tilde S$ and $\tilde c_V$ 
        in terms of their respective thermodynamic potentials $F$ and $U$.
        However, $\tilde c_V$ is no longer 
        related to $\tilde S$ through either type of temperature derivative.}
\label{fig5}
\end{figure}

In order to understand these other options, let us first recall 
the structure of the traditional thermodynamics. 
This structure is defined through the definitions in
Eq.~(\ref{usualrelations}), and is illustrated in Fig.~\ref{fig5}(a).
Note that the thermodynamic quantities are related to each 
other through temperature multiplications and differentiations,
forming a closed self-consistent set of definitions.
Of course, the temperature derivatives involved in these definitions
do not respect thermal duality, which is why we were motivated to construct
a thermal duality covariant temperature derivative.
Using this, we then developed a manifestly duality-covariant thermodynamics
by replacing all ordinary temperature derivatives with duality-covariant
derivatives.
This resulted in a version of thermodynamics
whose structure is illustrated in Fig.~5(b).
Indeed, as evident in Eq.~(\ref{newrelations}), 
our new quantities $\tilde S$ and $\tilde c_V$ are defined as 
covariant derivatives of their respective thermodynamic potentials $F$ and $U$.

%================== FIGURE ============================================
%================== END OF INSERTED FIGURE ============================

However, this replacement of $d/dT$ by $D_T$ 
does not preserve the entire structure of the traditional thermodynamics:
the final direct ``link'' between the entropy and specific heat is broken.
In the traditional thermodynamics, these two quantities are related by the identity
\beq
                     c_V ~=~ T {d\over dT} S~,
\label{oldlink}
\eeq
yet $\tilde c_V$ and $\tilde S$ are not related in this way
through either $T d/dT$ or $T D_T$.  (Note that since $\tilde S$ has zero weight,
$d/dT$ and $D_T$ are actually the same operator when acting on $\tilde S$.) 
Indeed, the fact that $\tilde c_V\not= T d \tilde S/dT$ is immediately
apparent upon comparing Figs.~\ref{fig1}, \ref{fig2}, and \ref{fig4}.

Of course, one might argue that preserving Eq.~(\ref{oldlink}) is not as
critical as preserving the identifications of the entropy and specific
heat as derivatives of their respective potentials.  However, 
in traditional thermodynamics,
the identity (\ref{oldlink}) is critical for interpreting
entropy in terms of heat transfer,
\beq
           dS~=~ {dQ\over T}~.
\label{entropyheat}
\eeq
To see this, recall that a heat transfer $dQ$ induces a change in internal energy $dU=dQ$
(where we are not distinguishing between exact and inexact differentials 
and where we have set $dW=0$).
However, since $dU= c_V dT$, we see that  
Eq.~(\ref{entropyheat}) cannot hold unless $S$ and $c_V$ are related
through Eq.~(\ref{oldlink}).

\begin{figure}[t]
\centerline{
   \epsfxsize 2.6 truein \epsfbox {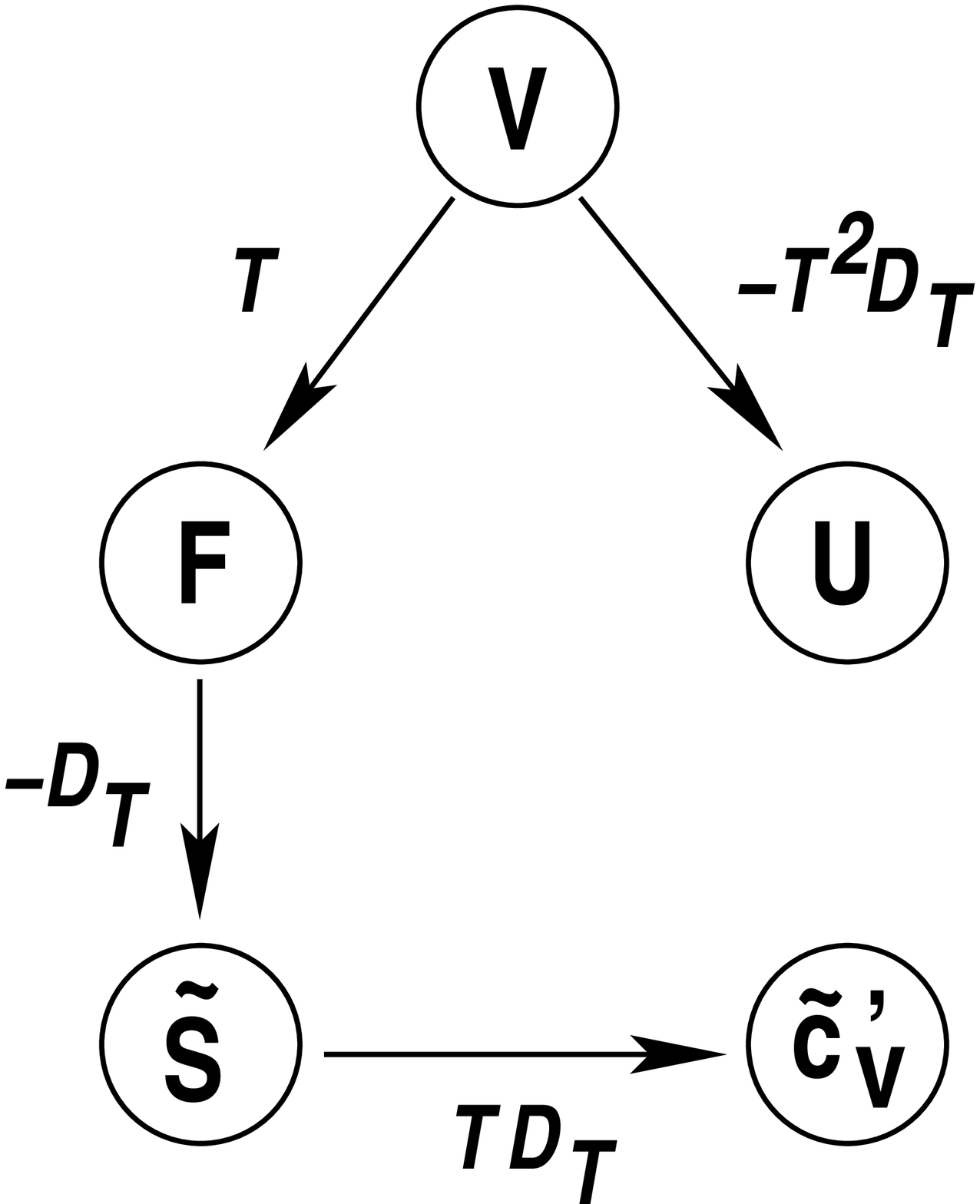}
   \hfill
   \epsfxsize 2.6  truein \epsfbox {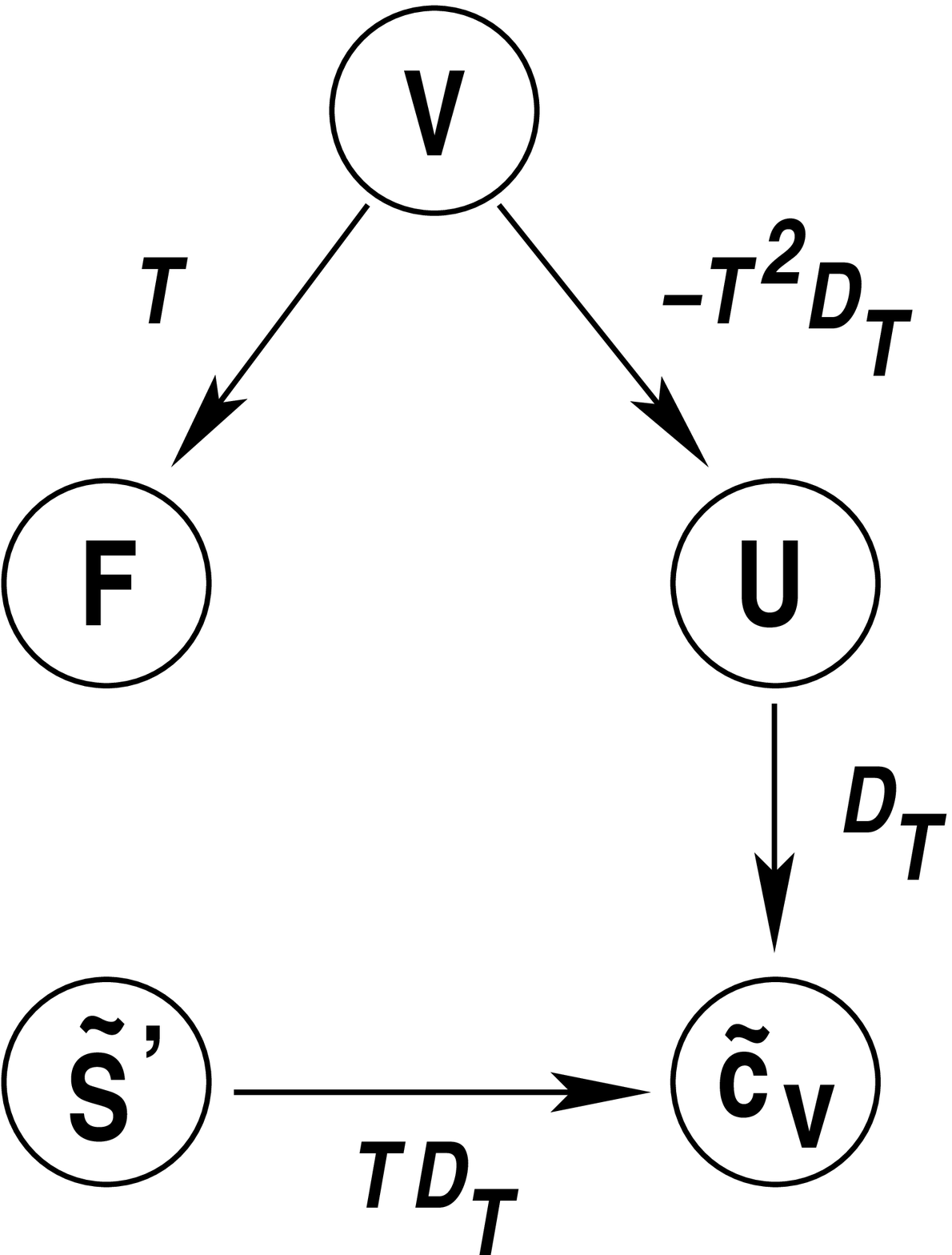}
    }
\caption{ 
   Relations between thermodynamic quantities in alternative formulations
    of duality-covariant thermodynamics. 
    (a) In this version based on Eq.~(\protect\ref{newlink1}), 
        the corrected entropy $\tilde S$ is defined
        through the free energy, but the corrected specific heat $\tilde c_V$
        is defined through the corrected entropy.
    (b) In this version based on Eq.~(\protect\ref{newlink2}), 
        the corrected specific heat is defined through
        the internal energy, and the corrected entropy is defined
        implicitly through the corrected specific heat.  }
\label{fig6}
\end{figure}

There are various ways in which this situation can be addressed.
One option, of course, is to regard the relation (\ref{oldlink}) as
more fundamental than the separate relations between
either the entropy or specific heat and their respective thermodynamic
potentials.  We could then establish a covariant thermodynamics by
replacing our previous definition for the corrected specific heat
with a new definition stemming directly from the corrected entropy:
\beq
            \tilde S ~=~ -D_T F~,~~~~~~~~~\tilde c^{\,\prime}_V ~=~ T D_T \tilde S~.
\label{newlink1}
\eeq
This option is illustrated in Fig.~\ref{fig6}(a).
Alternatively, we could retain the previous corrected specific heat $\tilde c_V$,
and implicitly define a new corrected entropy (up to an overall additive constant)
relative to this specific
heat:
\beq
           \tilde c_V  ~=~ D_T U~,~~~~~~~~~\tilde c_V ~=~ T D_T \tilde S'~.
\label{newlink2}
\eeq
This option is illustrated in Fig.~\ref{fig6}(b).

%================== FIGURE ============================================
%================== END OF INSERTED FIGURE ============================

Despite their differences,
each of these options results in a fully consistent, duality 
covariant thermodynamics.
The primary difference between them, of course, 
is in the interpretation given to their corrected entropies.
The corrected entropy $\tilde S$, which appears in Eq.~(\ref{newrelations})
and Eq.~(\ref{newlink1}), is derived from the free energy
which in turn is derived directly from the partition function $\calV$.
This entropy should thus retain its interpretation as a counting
of states (\ie, as a measure of disorder).  
The corrected entropy $\tilde S'$, by contrast,
is defined implicitly through Eq.~(\ref{newlink2}).  This entropy
should thus retain its interpretation pertaining to heat transfer.

Given these observations, the question then arises as to whether there exist
any special covariant derivatives $D_T$ for which {\it all}\/ of the ``links''
in these diagrams are generalized and continue to hold.
As we shall now prove, only one such derivative exists.

To see this, we first observe that the diagram in Fig.~\ref{fig5}(a)  
``closes'' for the usual thermodynamics as a result of the operator identity
\beq
          T {d^2\over dT^2} T  ~=~ {d\over dT} T^2 {d\over dT}~. 
\label{identity}
\eeq
This in turn holds as a result of the commutation relation 
\beq
               [T,d/dT] ~ =~ -1~.
\label{commrel1}
\eeq
Indeed, when acting on $\calV$, each side of Eq.~(\ref{identity}) 
provides a different route to the second derivative $c_V$:  the left side passes
through $F$ and $S$, while the right side passes through $U$.

We now seek to duplicate this success for our covariant derivative
$D_T$.  It is straightforward to demonstrate that 
\beq
          [T,D_T^{(k)}]~=~ -1~~~~~~~~~~ {\rm for~all}~~k~
\label{commrel2}
\eeq
where $k$ is the weight coefficient within $D_T$;
indeed, Eq.~(\ref{commrel1}) is nothing but the
$k=0$ special case of Eq.~(\ref{commrel2}).
However, in order to have our diagrams ``close'' for arbitrary
vacuum amplitudes $\calV$, our covariant derivatives must now satisfy
the generalized relation
\beq
           T D_T^{(0)} D_T^{(2)} T ~=~ D_T^{(2)} T^2 D_T^{(0)}~. 
\label{newidentity}
\eeq
Without loss of generality, let us write  $D_T^{(0)}=d/dT$ and 
$D_T^{(2)}= d/dT + g(T)/T$, as in Eq.~(\ref{genform}).
We then find
\beq
           [ D_T^{(2)}, D_T^{(0)} ] ~=~ 
                    - {d\over dT} \left( {g\over T}\right)~,
\eeq
which, along with the commutation relation in Eq.~(\ref{commrel2}), 
enables us to reduce Eq.~(\ref{newidentity})
to the differential equation
$dh/dT= -h/T$ where $h\equiv g/T$.
The only solution to this equation has $h\sim T^{-1}$, or $g$ equal to
a constant.\footnote{
     This conclusion can also be reached directly by observing that $\tilde c_V$ and $\tilde S$
     are related through the modified identity $\tilde c_V= T {d\tilde S/dT} + (dg/dT)F$.
     This reduces to the traditional identity only if $g(T)$ is a constant. 
     However, the above derivation involving the commutation relations
     of our covariant derivatives exposes the underlying algebraic structure 
     behind the failure of the traditional identity when $g(T)$ is not a constant.}
However, according to Eq.~(\ref{constr}), this constant must be equal
to $-k/2$. 
We thus find that $g(T)= -k/2$ is the unique solution 
which preserves
all of our thermodynamic identities, resulting in the unique covariant derivative
\beq
             D_T ~=~ {d\over dT} ~-~ {k\over 2T}~.
\eeq

Remarkably, this is precisely the derivative that we already found 
in Eq.~(\ref{Ddefzero}).
However, we see that we would now have to take this as our covariant
derivative for {\it all}\/ values of $T$ in order to preserve all of the 
``links'' in our covariant thermodynamics.
In other words, following Eq.~(\ref{shifts}),
we would have to define $\tilde S \equiv U/T$ for all $T$.
 
It is easy to see how this corrected entropy manages to retain both of its
interpretations pertaining to heat transfer and state counting.
By defining $\tilde S\equiv U/T$, we are providing a direct relation
between the corrected entropy and the internal energy, which in turn can be directly
related to heat.  A similar argument applies to counting states.
Recall that the traditional entropy $S$ is special in that it depends on the temperature
only though
the normalized Boltzmann probabilities $P_i \equiv p_i/Z$
where $p_i \equiv \exp(-E_i/T)$ and $Z\equiv \sum_i p_i$:
\beq
           S~=~ -\sum_i \, P_i \,\ln\, P_i~.
\label{usualentropyP}
\eeq
It is this expression which enables us to associate the emergence of
order with the vanishing of $S$:
as $T\to 0$, we find $P_i=0$ for all excited states and $P_i=1$ for the ground state.
However, if we now take $\tilde S\equiv U/T = S+F/T$, we find
\beqn
    \tilde S    ~=~  S + F/T &=& -\sum_i P_i \ln\, P_i  - \ln  Z\nonumber\\
                  &=&  -\sum_i P_i \left( \ln\, p_i - \ln\,Z \right) -  \ln Z\nonumber\\
                  &=& -\sum_i \, P_i \,\ln\, p_i~
\label{newentropyp}
\eeqn
where in the first line we have identified $F= -T\ln Z$ 
(as appropriate for the usual canonical ensemble).
We thus see that $\tilde S\equiv U/T$ is given by an expression which
is similar to 
Eq.~(\ref{usualentropyP})
but in which the final normalized Boltzmann probability $P_i$ is simply replaced
by the {\it unnormalized}\/  Boltzmann probability $p_i$.
Indeed, with this definition, $\tilde S$ is sensitive to the
distribution of {\it individual}\/ Boltzmann probabilities in precisely
the same way as the usual entropy $S$, and differs only in its dependence
on their combined sum.  Thus $\tilde S$ can also be taken as a direct
measure of disorder.

Unfortunately, these definitions fail a crucial test:  they do not
have a smooth limit as $T/T_c\to 0$ in which traditional thermodynamics is
restored.  Rather, this solution for $D_T$ exists only as a special point,
a unique alternative thermodynamics which does not connect smoothly
back to the traditional case.
We are therefore forced to disregard this possibility.
We see, then, that it is not generally possible to 
construct a thermal duality covariant thermodynamics which simultaneously
preserves {\it all}\/ of the traditional relations between our thermodynamic 
quantities.

%======================================================================================
\bigskip
\smallskip

\setcounter{footnote}{0}
\section*{Interlude\footnote{
    With apologies to Galileo, 
    whose {\it Dialogue on the Two Chief World 
    Systems}\/~\cite{Galileo}
     contains three characters:
    Simplicio (the conservative Aristotelian), 
    Salviati (the modernist), 
    and Sagredo (a neutral observer 
    who functions as commentator).}  -----------------------------------------\\} 

 %   FOR THE CURIOUS READER OF THIS RAW TeX FILE:  Concerning the Dialogues:
 %  The "modernist" is called Salviati, presumably named
 %  after Filipo Salviati, who until his death in 1614 had been a close friend of Galileo's in
 %  Florence. The clear-thinking representative of the educated public is called Sagredo, after
 %  Galileo's former student Giovanfrancesco Sagredo of Venice, who had died in 1620. The
 %  pedantic Aristotelian of the trio is called Simplicio, apparently named after Simplicius, a
 %  Greek philosopher of the 6th century AD known for his commentaries on Aristotle.
 %   NOW YOU KNOW.

\begin{itemize}

\item[{\bf Simplicio:}]  This is all very interesting, but I am still troubled by
    this proposal to modify traditional thermodynamics.  Even in condensed-matter 
    physics, there are systems (such as arrays of Josephson junctions) which 
    exhibit temperature-inversion symmetries which are analogues of thermal duality.  
    We do not modify the laws of thermodynamics when analyzing these
    systems;  we merely accept the fact that their free energies and their 
    entropies may exhibit the underlying symmetry to different degrees.
    Should the laws of thermodynamics really vary with the system?
    Should not the laws of thermodynamics {\it transcend}\/ the system under study? 

\item[{\bf Salviati:}]  You raise a good point.  However, this proposal
    is not about a particular system or configuration of matter.  Rather,
    this is a proposal whose inspiration is string theory, 
    a purported theory of matter itself at the most fundamental energy scales.
    Thus, the proposed string-theoretic corrections are to be interpreted
    as universal, valid for all systems regardless of their underlying symmetries.
    The analogy with gauge invariance is apt.  Classical electromagnetism
    is the theory underlying all electromagnetic phenomena, and it exhibits
    gauge invariance at its most fundamental level.  Within the framework of 
    this theory, regardless of the particular system or charge distribution under 
    study, we do not ascribe physical reality to quantities which are not gauge 
    invariant;  likewise we 
    would not tolerate a calculational methodology which explicitly breaks
    gauge invariance in a way that does not lead to gauge-invariant results.
    If thermal duality is truly a fundamental string symmetry,
    then the same should be true here.  Just as gauge invariance is used
    as a guide when performing calculations and extending our models into new
    domains, thermal duality is similarly being exploited to determine
    the forms of possible string-theoretic corrections 
    to the laws of thermodynamics.  We know that such corrections are necessary
    because the traditional laws
    explicitly break a symmetry which we are holding to be fundamental.

\item[{\bf Simplicio:}]  But the definition of entropy in standard thermodynamics
    preserves gauge invariance, as it should.  The trace involved in the
    partition function leads to a gauge-invariant free energy and also a 
    gauge-invariant entropy.

\item[{\bf Salviati:}]  Exactly.  But the critical difference here is that
     the thermal duality symmetry is one which involves temperature directly.  
     This is why the temperature derivative fails to commute with the symmetry,
     and requires covariantization. 

\item[{\bf Simplicio:}]  But must we take thermal duality as a fundamental symmetry?
    After all, thermal duality might simply be an accident of certain compactifications.

\item[{\bf Salviati:}]  This could indeed be the case.  However, thermal duality 
     is intimately related to T-duality and Lorentz
     invariance, and both of these are certainly fundamental symmetries in string theory.
     Indeed, T-duality is often taken as evidence that strings ``feel'' 
     the spacetime in which they propagate in a way that does not
     distinguish between large and small.  Symmetries such as these are not 
     considered accidents;  rather, they are taken as clues, evidence for the need
     to reinterpret the nature of time and space at the string scale.
     Since the roots of thermal duality are firmly embedded in T-duality,
     it would seem that the implications of thermal duality should be taken just 
     as seriously.  Thus, if T-duality tells us that our understanding of space itself  
     may require modification at the string scale, the correspondence between
     compactified zero-temperature theories and uncompactified finite-temperature
     theories suggests that the same must be true of our understanding of 
     thermodynamics.  It is then completely natural that the laws of thermodynamics
     would require modification.
     
\item[{\bf Simplicio:}]  But are there not closed string compactifications which fail to 
     exhibit thermal duality?
    
\item[{\bf Salviati:}]  Yes --- just as there are closed string compactifications which
     fail to be self-dual under T-duality transformations.
     Indeed, the analogy is exact at a mathematical level:  such compactifications
     have certain orbifold twists which mix into the compactification and spontaneously
     break the underlying symmetry.  However, the important point is that these
     are only {\it spontaneous}\/ breakings of the fundamental symmetry;  
     as with all Scherk-Schwarz breakings,
     the symmetry-breaking effects scale with the inverse volume of 
     the compactification and disappear in the infinite-volume limit.  
     The existence of compactifications
     in which these symmetries are spontaneously broken 
     does not alter the primary point that these are still fundamental symmetries
     in string theory, and we should not expect the {\it rules}\/ of the theory itself
     to violate them.  
     As stated in the Introduction, it is acceptable if the entropy $S(T)$ turns out 
     to be non-covariant because the underlying vacuum amplitude $\calV(T)$
     is non-covariant for a particular twisted string ground state.
     It is not acceptable, however, if the covariance of $S(T)$ is lost only because
     this quantity is defined in a way that 
     fails to respect the underlying symmetry.
     
\item[{\bf Simplicio:}]  So how, then, are we to interpret these string-corrected quantities
    $\tilde S$ and $\tilde c_V$?  It seems that the most conservative approach would be 
    to assert that these new quantities are merely the proper ``eigenquantities'' with 
    respect to thermal duality transformations, and that the entropy and specific
    heat are not eigenquantities but rather linear combinations of these eigenquantities.
    This alone would be very interesting.  Why not rest there?  Why is it necessary
    to impose the further interpretation that $\tilde S$ is itself the actual entropy,
    that $\tilde c_V$ is itself the actual specific heat?  In other words, must we interpret
    the extra string-suppressed terms in the definitions of $\tilde S$ and $\tilde c_V$
    as {\it corrections}\/?
    
\item[{\bf Salviati:}]  One could indeed adopt the conservative interpretation 
    you are proposing.  However, we would then be placed in the somewhat awkward 
    position of associating physical observables such as entropy and specific heat with
    mathematical quantities that fail to exhibit our fundamental symmetries.
    If we believe fully that entropy and specific heat are physical observables,
    we are motivated to associate them with mathematical 
    quantities such as $\tilde S$ and $\tilde c_V$ which are consistent with
    these symmetries.  This is indeed a more ambitious interpretation of the results, but
    this seems especially natural in light of the fact that the extra terms involved
    are, as noted, suppressed by powers of the string scale and hence are unobservable
    at low temperatures.

\item[{\bf Simplicio:}]  This still seems troubling.  According to the ``strong''
    interpretation you are advancing, a quantity such as entropy now has
    an extra contribution in its definition, one which depends on an energy
    scale $T_c$ which is in turn related to the string scale.  
    How can this be justified, given our expectation that entropy is merely
    a counting of states?  Should not entropy be a pure number without
    reference to any physical scale?  This goes back to my original question:
    should not the formulation of thermodynamics {\it transcend}\/   
    a particular theory?
 
\item[{\bf Salviati:}]  Yes, I understand your concern.  
    I shall propose two answers.  
    First, the covariant derivative with $\alpha=0$ actually
    does not introduce any new scale $T_c$.  Moreover, 
    this is the unique derivative which restores thermal duality while 
    simultaneously managing to close all of the ``links''
    in the thermodynamics diagrams in Sect.~7. 
    Of course, this derivative does not admit traditional thermodynamics
    as a low-temperature limit, thus requiring that it be interpreted only
    as strongly as your ``conservative'' approach would permit. 
    The second answer, however, goes perhaps more to the point you are raising.
    It is certainly true that one is, in general, introducing a 
    physical scale into the definition of entropy;  this was hardly
    to be avoided, since the symmetry one is attempting to restore
    by doing so also contains a physical scale.  However, this is not just any
    scale:  this is the fundamental scale of string theory, the scale
    which one expects to govern the relative sizes of string-related
    phenomena associated with quantum gravity and a possible breakdown
    of our usual notions of spacetime geometry.
    It is not too much
    to imagine that this profound alteration should also affect 
    the very meaning of degrees of freedom and counting of states;  
    indeed, one suspects a connection with holography here.

\item[{\bf Simplicio:}]  But this is precisely my worry.
    I could certainly understand if the number of degrees of freedom
    in the theory changes as one approaches the string scale due
    to some hitherto unknown gravitational or string-induced effect.
    This would indeed be in the spirit of holography.   
    However, the proposed modification of the laws of thermodynamics
    does not appear to be changing physical quantities such as the 
    degrees of freedom of the theory;  by redefining entropy, it 
    merely changes the probabilistic rules by which they are counted.  
    Surely this cannot be correct.

\item[{\bf Salviati:}]  Ah, but it may.  Even with the usual definition of
    entropy, we count all states equally because we assume that each microstate
    of the system is equally likely to occur, that a given system explores
    all of its energetically allowed states with equal probability.  
    This assumption
    is ultimately the bedrock of standard thermodynamics, but it is
    possible that this assumption is violated at the string scale.
    After all, we already know that this assumption is violated in
    purely classical (deterministic) systems, which must 
    obey the Poincar\'e recurrence theorem and hence cannot truly
    explore the space of states completely randomly.
    In such systems, the validity of such an assumption becomes a question
    of timescales, and these ultimately depend on the 
    relevant physical parameters of the system. 
    Even in a quantum-mechanical system, this assumption is justified
    only in a rough statistical sense, thanks to quantum-mechanical
    uncertainties in specifying our states;  once again, the validity
    of the assumption depends on the physical parameters of the system. 
    It is therefore not too much to expect that near the string scale,
    new quantum-gravitational or string-induced effects may 
    also ultimately distort the manner in which the system explores all of 
    its energetically allowed states.
    If so, the string-corrected entropy may be precisely what accounts for
    this phenomenon, providing a recipe for computing an ``effective'' 
    number of degrees of freedom after all gravitational or string-induced
    effects are included.  Indeed, as long as the final corrected entropy  
    exhibits thermal duality along with the other thermodynamic quantities, 
    who can say whether the true change is in the number of degrees of
    freedom or in the manner by which they are counted?  Only
    the final count is important.

\item[{\bf Sagredo:}]  Gentlemen, gentlemen, I fear your discussion is
    becoming too philosophical for my tastes.  We all admit that the
    proposed thermodynamics differs
    from the standard thermodynamics only through effects which
    are unmeasurably small at temperatures much below the string scale.  
    Given that physics is an experimental science, we cannot 
    prove or disprove this proposal except through recourse to aesthetics.  
    In this case, aesthetics means symmetry.  The proposed modifications
    to thermodynamics restore one symmetry, namely thermal duality,
    but imply profound changes to our understanding of entropy
    which make our dear friend Simplicio very nervous.  Perhaps we should
    have expected that our understanding of quantities such as entropy
    might require profound alteration as we approach the fundamental scale
    of quantum gravity and string theory.  In any case, I fear this discussion
    is best continued in private, perhaps lubricated with a glass of
    foamy ale. 

\item[{\bf Simplicio:}]  Agreed.  But, as we are all Renaissance Italians, perhaps 
    a fine wine would be more appropriate...

\end{itemize}

\bigskip

%==================================================================================
\section{Conclusions and Open Questions}
\setcounter{footnote}{0}

In this paper, we have addressed a fundamental issue:
is it possible to construct a thermodynamics
which is manifestly covariant with respect to the thermal duality
symmetry of string theory?

In one sense, this approach was successful.
We were able to construct a manifestly covariant derivative, and
through this derivative we were able to constuct a manifestly
covariant thermodynamics which not only 
reduces to the standard thermodynamics at low temperatures,
but which leads to corrections 
that become significant only near the string scale.
As pointed out by Sagredo, this alone guarantees that such
a theory is experimentally viable as an extension to the standard rules
of thermodynamics.  Given that this theory restores a fundamental
duality symmetry where it was otherwise lacking, we believe
that such extensions to the rules of thermodynamics   
are worthy of further exploration.

Adopting this attitude, we are then led to a number of outstanding
questions.  First, of course, there are several theoretical issues.
Most importantly, we needed to make an assumption
for the form of the function $g(T)$ in our covariant derivative.
While many of our main conclusions are independent of the specific
form of $g(T)$, it still remains to calculate this function from first
principles through a string calculation analogous
to that discussed in Sect.~3.  This would, we believe, place
our proposal on firmer theoretical footing. 
Another theoretical issue concerns the possible relation, if any,
between our results and holography.  Given that we are changing
the rules by which entropy is to be calculated --- indeed changing
the very definition of entropy itself --- it is important to study
whether and how the effects of these string corrections can
be interpreted in a holographic context.  We have already seen,
for example, that in many cases these string corrections tend to
profoundly alter the scaling behavior of the entropy with temperature,
thereby decreasing the effective spacetime dimensionality associated with
the entropy.  However, as discussed earlier,
interpreting this effect as truly ``holographic'' would also require
a geometric understanding of how the degrees of freedom 
contributing to $\tilde S$ may be mapped from a volume to the
boundary of a volume.  This issue cannot be addressed in our
formulation which is thus far based on strings in flat (infinite-volume)
backgrounds.

There are also many phenomenological issues that are prompted by our approach.
For example, how do our results extend to theories in which thermal
duality is spontaneously broken (see, \eg, 
Refs.~\cite{Rohm,AlvOsoNPB304,McGuigan,AtickWitten,KounnasRostand,shyam,III}), 
as well as to open strings and branes?
The answers to these questions could have important 
implications for recent brane-world scenarios.
Likewise, it is interesting to consider the possible applications of
our results to early-universe cosmology, particularly regarding
the issues of Hagedorn-like phase transitions and entropy generation.

In another sense, however, our investigations have perhaps raised
more questions than they have answered, and in this regard
we are quite sympathetic to the discomfort of Simplicio.
The structure of thermodynamics is so tightly constrained, 
and the underpinnings of thermodynamics rest on such elementary
axioms of probability and state-counting, that it would
seem to be an extremely risky undertaking to attempt any alteration
or generalization of these principles.  
We have already seen in Sect.~7, for example, that there are
several possible generalizations of the traditional rules of
thermodynamics, yet none of these approaches simultaneously 
preserves all of the different shades of interpretation that 
are normally ascribed to quantities such as entropy.  

Many of these theoretical issues could perhaps be resolved
(or at least placed on firmer footing) 
if we were to develop a formulation
of our generalized thermodynamics based on the {\it microcanonical}\/ 
ensemble. Yet we can immediately see the difficulties in doing
so.  By its very nature, thermal duality is a symmetry with respect
to transformations in temperature;  clearly temperature is
the independent variable.  In order to develop an equivalent microcanonical
formulation, however, we require the internal energy $U$ to be
the independent variable.
We would thus need to express thermal duality as a symmetry
under transformations of $U$. 
We would then attempt to take our string-corrected entropy $\tilde S$ 
as the fundamental quantity (\ie, the string-corrected counting of states),
and demonstrate that $d\tilde S/dU$ (or even a covariant
derivative $D_U \tilde S$) is equivalent to the inverse of our original 
temperature $1/T$.  However, it is easy to verify that this 
microcanonical approach
does not generally lead
to results which are consistent with those of the canonical ensemble.  
Indeed, we believe that the fundamental difficulty 
in this approach rests on the need to find a microcanonical-ensemble 
equivalent of thermal duality --- \ie, a formulation
of this symmetry which does not take $T$ as the independent parameter.
As long as our approach to string thermodynamics rests
on the canonical ensemble and string partition functions, this
formulation is likely to elude us.
Similar issues concerning the relation between the microcanonical and canonical
ensembles are well known to exist in attempting to understand
the Hagedorn transition, and may also play a role in generic problems concerning
the interplay between gravity and thermodynamics, such as the Jeans instability.

Thus, echoing Simplicio, what are we to make of these results?
On the one hand, we could be content with the observation that there 
exist special solutions for $\calV(T)$, as discussed in Ref.~\cite{I},
for which the traditional entropy $S$ (and occasionally even the specific heat $c_V$)
turn out to be duality covariant.  Indeed, in Ref.~\cite{I}, we conjectured
that these special solutions $\calV(T)$
may represent the exact results of actual string calculations when the contributions
from all orders in string perturbation theory are included.  However, 
we continue to remain sympathetic to the original motivation of this paper, namely
that the rules of thermodynamics should themselves respect this symmetry in a manifest fashion.  
Indeed, it is by thrashing out how this can occur that we continue to hope to gain insight into
the possible nature of temperature, state counting, and thermodynamics near 
the string scale. 
After all, if thermal effects can truly be associated with 
spacetime compactification through the Matsubara/Kaluza-Klein correspondence,
then our expectations of an unusual ``quantum geometry'' near the string scale --- 
one which does not distinguish between ``large'' and ``small'' ---
should simultaneously lead to expectations of an equally unusual thermodynamics
near the string scale which does not distinguish between ``hot'' and ``cold''
in the traditional sense.  Thermal duality should then serve as a tool 
towards deducing the nature of these new effects.
We thus consider the investigation in this paper to be an initial, and hopefully
provocative, attempt in this direction.

%=========================================================================== 
\section*{Acknowledgments}
\setcounter{footnote}{0}

This work is supported in part by the National Science Foundation
under Grants~PHY-0071054 and PHY-0301998,
and by a Research Innovation Award from Research Corporation.
We wish to thank
S.~Chaudhuri, I.~Mocioiu, R.~Roiban, U.~van~Kolck, and especially
C.~Stafford for discussions.

\bigskip
\vfill\eject

%=================================================================================
\bibliographystyle{unsrt}

\begin{thebibliography}{99}



\bibitem{I}  K.R. Dienes and M.D. Lennek, 
    {\it Adventures in Thermal Duality (I):  Extracting Closed-Form Solutions
        for Finite-Temperature Effective Potentials in String Theory},
           arXiv:hep-th/0312216.
       %%CITATION = HEP-TH 0312216;%%

\bibitem{summary}  K.~R.~Dienes and M.~Lennek,
      {\it Thermal Duality Confronts Entropy:  A New Approach to String Thermodynamics?},
       arXiv:hep-th/0312173.
       %%CITATION = HEP-TH 0312173;%%

\bibitem{OBrienTan}
      K.~H.~O'Brien and C.~I.~Tan,
      %``Modular Invariance Of Thermopartition Function And Global Phase Structure Of Heterotic String,''
      Phys.\ Rev.\ D {\bf 36} (1987) 1184.
      %%CITATION = PHRVA,D36,1184;%%


\bibitem{AlvOsoNPB304}
       E.~Alvarez and M.~A.~R.~Osorio,
       %``Cosmological Constant Versus Free Energy For Heterotic Strings,''
       Nucl.\ Phys.\ B {\bf 304} (1988) 327
       [Erratum-ibid.\ B {\bf 309} (1988) 220].
       %%CITATION = NUPHA,B304,327;%%

\bibitem{AtickWitten}
           J.~J.~Atick and E.~Witten,
           %``The Hagedorn Transition And The Number Of Degrees Of Freedom Of String Theory,''
           Nucl.\ Phys.\ B {\bf 310} (1988) 291.
           %%CITATION = NUPHA,B310,291;%%


\bibitem{AlvOsoPRD40}
        E.~Alvarez and M.~A.~R.~Osorio,
       %``Duality Is An Exact Symmetry Of String Perturbation Theory,''
       Phys.\ Rev.\ D {\bf 40} (1989) 1150.
       %%CITATION = PHRVA,D40,1150;%%


\bibitem{OsoIJMP}
         M.~A.~R.~Osorio,
         %``Quantum fields versus strings at finite temperature,''
         Int.\ J.\ Mod.\ Phys.\ A {\bf 7} (1992) 4275.
         %%CITATION = IMPAE,A7,4275;%%

\bibitem{Polbook}
      For an introduction, see J. Polchinski, {\it String Theory, Vol.~I}\/
      (Cambridge University Press, 1998), Chap.~9.

\bibitem{Pol86}   J.~Polchinski,
     %``Evaluation Of The One Loop String Path Integral,''
     Commun.\ Math.\ Phys.\  {\bf 104} (1986) 37.
     %%CITATION = CMPHA,104,37;%%


\bibitem{SakaiSenda}
       N.~Sakai and I.~Senda,
       %``Vacuum Energies Of String Compactified On Torus,''
       Prog.\ Theor.\ Phys.\  {\bf 75} (1986) 692
       [Erratum-ibid.\  {\bf 77} (1987) 773].
       %%CITATION = PTPKA,75,692;%%


\bibitem{Nairetal}
       V.~P.~Nair, A.~D.~Shapere, A.~Strominger and F.~Wilczek,
       %``Compactification Of The Twisted Heterotic String,''
       Nucl.\ Phys.\ B {\bf 287} (1987) 402.
       %%CITATION = NUPHA,B287,402;%%


\bibitem{Sathiapalan}
       B.~Sathiapalan,
      %``Duality In Statistical Mechanics And String Theory,''
      Phys.\ Rev.\ Lett.\  {\bf 58} (1987) 1597.
      %%CITATION = PRLTA,58,1597;%%


\bibitem{McClainRoth}
      B.~McClain and B.~D.~B.~Roth,
      %``Modular Invariance For Interacting Bosonic Strings At Finite Temperature,''
      Commun.\ Math.\ Phys.\  {\bf 111} (1987) 539.
      %%CITATION = CMPHA,111,539;%%

\bibitem{review}  For a review, see
        K.~R.~Dienes,
        %``String Theory and the Path to Unification: A Review of Recent Developments,''
        Phys.\ Rept.\  {\bf 287} (1997) 447
        [arXiv:hep-th/9602045].
        %%CITATION = HEP-TH 9602045;%%

\bibitem{Kaplunovsky}  
           V.~S.~Kaplunovsky,
           %``One Loop Threshold Effects In String Unification,''
           Nucl.\ Phys.\ B {\bf 307} (1988) 145
           [Erratum-ibid.\ B {\bf 382} (1992) 436]
           [arXiv:hep-th/9205070].
           %%CITATION = HEP-TH 9205070;%%

\bibitem{KiritsisKounnas}  
       E.~Kiritsis and C.~Kounnas,
      %``Infrared regularization of superstring theory and the one loop calculation of coupling constants,''
      Nucl.\ Phys.\ B {\bf 442} (1995) 472
      [arXiv:hep-th/9501020].
      %%CITATION = HEP-TH 9501020;%%


\bibitem{Hagedorn}
        R.~Hagedorn,
        %``Statistical Thermodynamics Of Strong Interactions At High-Energies,''
        Nuovo Cim.\ Suppl.\  {\bf 3} (1965) 147.
        %%CITATION = NUCUA,3,147;%%
       
\bibitem{Huang}
       K.~Huang and S.~Weinberg,
       %``Ultimate Temperature And The Early Universe,''
       Phys.\ Rev.\ Lett.\  {\bf 25} (1970) 895.
       %%CITATION = PRLTA,25,895;%%

\bibitem{Bowick}
      M.~J.~Bowick and L.~C.~R.~Wijewardhana,
     %``Superstrings At High Temperature,''
     Phys.\ Rev.\ Lett.\  {\bf 54} (1985) 2485.
     %%CITATION = PRLTA,54,2485;%%

\bibitem{Tye}
       S.~H.~H.~Tye,
       %``The Limiting Temperature Universe And Superstring,''
       Phys.\ Lett.\ B {\bf 158} (1985) 388.
       %%CITATION = PHLTA,B158,388;%%

\bibitem{Alvarez}
         E.~Alvarez,
         %``Strings At Finite Temperature,''
         Nucl.\ Phys.\ B {\bf 269} (1986) 596.
         %%CITATION = NUPHA,B269,596;%%


\bibitem{Sathiapalan2}
        B.~Sathiapalan,
       %``Vortices On The String World Sheet And Constraints On Toral Compactification,''
       Phys.\ Rev.\ D {\bf 35} (1987) 3277;\\
       %%CITATION = PHRVA,D35,3277;%%
      Y.~I.~Kogan,
      %``Vortices On The World Sheet And String's Critical Dynamics,''
      JETP Lett.\  {\bf 45} (1987) 709
      [Pisma Zh.\ Eksp.\ Teor.\ Fiz.\  {\bf 45} (1987) 556].
      %%CITATION = JTPLA,45,709;%%

\bibitem{AlvOsoPRD36}
        E.~Alvarez and M.~A.~R.~Osorio,
        %``Superstrings At Finite Temperature,''
        Phys.\ Rev.\ D {\bf 36} (1987) 1175.
        %%CITATION = PHRVA,D36,1175;%%

\bibitem{LeBlanc}
       Y.~Leblanc,
       %``Cosmological Aspects Of The Heterotic String Above The Hagedorn Temperature,''
       Phys.\ Rev.\ D {\bf 38} (1988) 3087.
       %%CITATION = PHRVA,D38,3087;%%


\bibitem{Deo}
       N.~Deo, S.~Jain and C.~I.~Tan,
       %``Strings At High-Energy Densities And Complex Temperature,''
       Phys.\ Lett.\ B {\bf 220} (1989) 125;
       %%CITATION = PHLTA,B220,125;%%
       %``String Statistical Mechanics Above Hagedorn Energy Density,''
       Phys.\ Rev.\ D {\bf 40} (1989) 2626.
       %%CITATION = PHRVA,D40,2626;%%

\bibitem{BowickGiddings}
       M.~J.~Bowick and S.~B.~Giddings,
       %``High Temperature Strings,''
       Nucl.\ Phys.\ B {\bf 325} (1989) 631.
       %%CITATION = NUPHA,B325,631;%%
      
\bibitem{Giddings}
      S.~B.~Giddings,
      %``Strings At The Hagedorn Temperature,''
      Phys.\ Lett.\ B {\bf 226} (1989) 55.
      %%CITATION = PHLTA,B226,55;%%

\bibitem{AntonKounnas}
      I.~Antoniadis and C.~Kounnas,
      %``Superstring phase transition at high temperature,''
      Phys.\ Lett.\ B {\bf 261} (1991) 369.
      %%CITATION = PHLTA,B261,369;%%


\bibitem{Galileo}
      G.~Galilei, {\it Dialogue Concerning the Two Chief World Systems}\/ (Florence, 1632)
      [arXiv:astro-ph/3201001].
      %%CITATION = ASTRO-PH 3201001;%%


\bibitem{Rohm}
      R.~Rohm,
      %``Spontaneous Supersymmetry Breaking In Supersymmetric String Theories,''
      Nucl.\ Phys.\ B {\bf 237} (1984) 553;\\
      %%CITATION = NUPHA,B237,553;%%
     H.~Itoyama and T.~R.~Taylor,
      %``Supersymmetry Restoration In The Compactified O(16) X O(16)-Prime Heterotic
     %String Theory,''
     Phys.\ Lett.\ B {\bf 186} (1987) 129;\\
     %%CITATION = PHLTA,B186,129;%%
     P.~Ginsparg and C.~Vafa,
     %``Toroidal Compactification Of Nonsupersymmetric Heterotic Strings,''
     Nucl.\ Phys.\ B {\bf 289} (1987) 414;\\
     %%CITATION = NUPHA,B289,414;%%
     J.~D.~Blum and K.~R.~Dienes,
     %``Duality without supersymmetry: The case of the SO(16) x SO(16) string,''
     Phys.\ Lett.\ B {\bf 414} (1997) 260
     [arXiv:hep-th/9707148];
     %%CITATION = HEP-TH 9707148;%%
     %``Strong/weak coupling duality relations for non-supersymmetric string
     %theories,''
     Nucl.\ Phys.\ B {\bf 516} (1998) 83
     [arXiv:hep-th/9707160];
     %%CITATION = HEP-TH 9707160;%%
     %``From the type I string to M-theory: A continuous connection,''
     Nucl.\ Phys.\ B {\bf 520} (1998) 93
     [arXiv:hep-th/9708016].
     %%CITATION = HEP-TH 9708016;%%    


\bibitem{McGuigan}
     M.~McGuigan,
     %``Finite Temperature String Theory And Twisted Tori,''
     Phys.\ Rev.\ D {\bf 38} (1988) 552.
     %%CITATION = PHRVA,D38,552;%%


\bibitem{KounnasRostand}
      C.~Kounnas and B.~Rostand,
      %``Coordinate Dependent Compactifications And Discrete Symmetries,''
      Nucl.\ Phys.\ B {\bf 341} (1990) 641.
      %%CITATION = NUPHA,B341,641;%%

\bibitem{shyam}
     S.~Chaudhuri,
      %``Finite temperature bosonic closed strings: Thermal duality and the KT
     %transition,''
     Phys.\ Rev.\ D {\bf 65} (2002) 066008
     [arXiv:hep-th/0105110];
     %%CITATION = HEP-TH 0105110;%%
     %``Finite temperature gases of fermionic strings,''
     arXiv:hep-th/0208112.
     %%CITATION = HEP-TH 0208112;%%


\bibitem{III}
      K.~R.~Dienes and M.~Lennek, in preparation.



\end{thebibliography}

\end{document}